\documentclass{aastex63}

\usepackage{amsmath}

\received{February 5, 2021}
\revised{April 26, 2021}
\accepted{April 26, 2021}

\submitjournal{ApJ}

\shorttitle{BD Binary Clouds}
\shortauthors{Brock et al.}

\newcommand{\microns}{$\mu$m}
\newcommand{\PHOENIX}{\tt PHOENIX}

\graphicspath{{./}{figures/}}

\begin{document}

\title{Cloud Properties of Brown Dwarf Binaries Across the L/T Transition}

\correspondingauthor{Laci S. Brock}
\email{laci@lpl.arizona.edu}

\author[0000-0002-0786-7307]{Laci Shea Brock}
\affiliation{Lunar and Planetary Laboratory
1629 E. University Blvd.,
Tucson, AZ 85721 USA}

\author[0000-0002-7129-3002]{Travis Barman}
\affiliation{Lunar and Planetary Laboratory
1629 E. University Blvd.,
Tucson, AZ 85721 USA}

\author[0000-0002-9936-6285]{Quinn M. Konopacky}
\affiliation{Center for Astrophysics and Space Sciences
9500 Gilman Dr.,
La Jolla, CA 92093 USA}

\author[0000-0003-0454-3718]{Jordan M. Stone}
\affiliation{Naval Research Laboratory,
Remote Sensing Division,
4555 Overlook Ave SW,
Washington, DC 20375 USA}

\begin{abstract}
We present a new suite of atmosphere models with flexible cloud parameters to investigate
the effects of clouds on brown dwarfs across the L/T transition.  We fit these models to a
sample of 13 objects with well-known masses, distances, and spectral types spanning L3-T5.
Our modelling is guided by spatially-resolved photometry from the Hubble Space Telescope
and the W. M. Keck Telescopes covering visible to near-infrared wavelengths.  We find
that, with appropriate cloud parameters, the data can be fit well by atmospheric models
with temperature and surface gravity in agreement with the predictions of
evolutionary models.
We see a clear trend in the cloud parameters with spectral type, with earlier-type objects
exhibiting higher-altitude clouds with smaller grains (0.25-0.50 $\micron$) and later-type
objects being better fit with deeper clouds and larger grains ($\geq$1 $\micron$). Our
results confirm previous work that suggests L dwarfs are dominated by submicron particles,
whereas T dwarfs have larger particle sizes.\end{abstract}

\keywords{brown dwarfs: binaries, atmospheres, clouds}

\section{Introduction}
As important opacity sources, clouds play a major role in determining the atmospheric
structures, emergent spectra, and evolution of brown dwarfs.  Clouds fundamentally impact
the observational properties of color, magnitude, and spectra with compositions that span
a diverse mix of condensates dependent upon a wide range of effective temperatures,
gravities, metallicities, and pressures \citep{Helling2014,Marley2015}.  The transition
from L to T spectral types is perhaps the most well-known impact of changes in cloud
opacity with observations of over 1000 field dwarfs highlighting the drastic shift in
near-infrared colors from red to blue that occurs around a narrow range of effective
temperatures ($\approx$1400 K) \citep{Faherty2012,Dupuy2012,Leggett2002,Kirkpatrick2005}.

Atmospheres of L dwarfs are dominated by refractory materials condensing and forming optically thick cloud layers \citep[e.g.,][]{Tsuji1996,Allard2001,Cushing2008, Marley2002}. The common species include iron (Fe), corundum (Al$_{2}$O$_{3}$), enstatite (MgSiO$_{3}$), and forsterite (Mg$_{2}$SiO$_{4}$). T dwarfs mark the disappearance of thick iron and silicate clouds from the photosphere and the appearance of methane absorption \citep{Kirkpatrick2005,Marley2010,Burgasser2003}. Matching observed colors and spectra of L dwarfs requires some type of cloud layer in atmosphere models \citep{Burrows2006}. Mid-T dwarfs and cooler types are often best approximated using cloud-free atmosphere models \citep{Kirkpatrick2005}; however, in some cases models with optically thin or inhomogeneous clouds composed of low temperature condensates (Na$_{2}$S, KCl, ZnS, MnS, Cr) can match T-dwarf photometry \citep{Morley2012,Charnay2018}.

Self-consistent, one-dimensional atmosphere models have been the most widely used to study
brown dwarfs.  While they do not capture the atmospheric dynamics
\citep[e.g.,][]{Showman2019,Zhang2014,Tan2019,Tan2021,Tan2021b,Charnay2018} they do an excellent
job of capturing the average properties and allow one to study the spectral trends across the brown
dwarf population with a homogeneous set of modelling assumptions.  The most recent model spectra
have reached a high degree of realism through the development of molecular opacity
databases \citep{Freedman2008,Sharp2007,Tennyson2012}, the chemistry of gas and condensate
species \citep{Lodders2002}, and resonance line broadening \citep{Burrows2001,Allard2001}.
Incorporating parameterized clouds into atmosphere models better facilitates the
interpretation of observational data, and these approaches have shown that L dwarfs and
early T dwarfs are better represented by cloudy models rather than previous cloud-free
models \citep{Marley2002,Ackerman2001,Allard2001,Tsuji1996}. In general, if appropriate
input is provided, synthetic photometry, spectra, and colors generated by models are able
to reproduce observational data quite well \citep{Saumon2008}.

Cloudy and cloud-free limits provide useful insight into the photometric and spectroscopic
trends across brown dwarf spectral types, but these limiting cases do not accurately
predict the scatter of colors within the L spectral type \citep{Marley2010} and often
fail to match individual brown dwarfs \citep{Burrows2006}. \citet{Liu2016} confirmed that low-gravity objects occupy a distinct region of the color-magnitude diagram separate from field brown dwarfs of the same spectral type. The complexities pose additional challenges in understanding how brown dwarfs evolve through time, especially across the L/T transition($\approx$ L8-T4), where major changes in atmospheric chemistry and dynamics occur. L dwarfs with red near-infrared colors are associated with signs of youth \citep{Kirkpatrick2008,Cruz2009}, low gravity, and/or high metallicity \citep{Mclean2003,Looper2008b,Stephens2009}; whereas bluer L dwarfs are typically associated with an older age, higher gravity, and/or low metallicity \citep{Schmidt2010,Burgasser2010,Cushing2010}. Secondary influences from clouds can cause additional color dispersion, ranging from the thickness of the clouds to larger grain sizes \citep{Knapp2004, Burgasser2008a}. Although models can reproduce spectra of red L dwarfs with thick condensate clouds, there has been disagreement in effective temperatures across models \citep{Gizis2012}. The challenge is developing comprehensive atmosphere models while disentangling the effects of local cloud properties (e.g., thickness, grain size) within an atmosphere from global parameters (e.g., surface gravity, metallicity). Recent advancements use rotational modulation coupled with heterogeneous cloud models that include disequilibrium CO/CH$_{4}$ chemistry to isolate differences between cloud-induced and gravity-induced features \citep{Lew2016,Lew2020}.

Another approach to this problem is to investigate the properties of clouds in a sample of well-studied brown dwarf binary systems with precisely measured properties. Resolved photometry and spectroscopy for field age binaries of known distance and mass
provide the clearest picture of the L-T spectral sequence as coevality can be assumed for
such systems. The two components often have similar compositions, surface gravities,
relatively constant radii for their age ($\geq$ 0.5 Gyr), and tend to have nearly equal
masses \citep{Dupuy2012}. Furthermore, individually resolved brown dwarf binaires provide the most robust tests for atmospheric and evolutionary models to-date through precisely measured dynamical masses.  Mass is one of the most fundamental parameters of brown dwarfs that can aid in unravelling atmospheric complexities through its influence on surface gravity and
evolution; however, mass is often difficult to measure and observations are complicated by
the largely unknown span of ages for field objects due to degeneracies between mass and age
\citep{Konopacky2010,Dupuy2017}.

Testing and constraining substellar evolutionary models requires benchmark brown dwarfs--objects with independently determined masses, luminosities, and ages. Early work hinted at a luminosity problem where comparisons between dynamical masses and masses predicted by evolutionary models differed significantly \citep{Dupuy2009a,Konopacky2010,Dupuy2014}. The consistency between predicted and measured masses has improved with continued astrometric monitoring \citep{Dupuy2017} and models that consider cloud clearing \citep{Saumon2008,Allard2013}. A related problem exists for benchmark brown dwarfs when comparing the effective temperatures and gravities predicted by evolution models to those obtained by comparing model spectra to observed photometry or spectra. Differences in temperature as large as several hundred Kelvin and differences in gravity of an order of magnitude are not uncommon across the L/T transition \citep{Cushing2008}. Clouds can alleviate some of the discrepancy between evolutionary and atmosphere model comparisons but does not eliminate the problem completely \citep{Wood2019}. Some of the issues may be improved with very broad wavelength coverage \citep{Briesemeister2019}. In addition to refined cloud parameters, metallicity \citep{Wood2019,Crepp2018}, non-equilibrium chemistry \citep{Barman2011,Zhang2020}, or other missing physics in the models \citep{Charnay2018} (e.g., cloud microphysics) may be remaining sources of tension.

The importance of clouds shaping observed properties across the L/T transition motivates our
work to investigate cloud properties of individual brown dwarf binaries.  An increase in
the number of binary systems with known distance and mass across a broad range of
temperatures allow us to explore the validity of previous discrepancies between predictions
of atmosphere and evolutionary models. 

In this paper, we test whether cloudy atmosphere models can be produced that match
multi-band photometric observations to a reasonable degree.  We focus on a sample of seven
brown dwarf binary systems that span mid-L to mid-T in spectral type.  By using a small
sample size and a range of spectral types on either side of the L/T transition, we are
able to explore cloud properties in finer detail across a broad range of temperature and
gravity.  In addition to studying the cloud properties, we determine the effective
temperatures for each binary component in the sample, independent of spectral type. 
Near-infrared photometry can be used to estimate T$_{\mathrm{eff}}$; however, because
T$_{\mathrm{eff}}$ is a bolometric quantity, broad coverage of the spectral energy
distribution (SED), especially on both sides of $\lambda_{\mathrm{max}}$, is highly
desirable.  We are able to determine a robust value of T$_{\mathrm{eff}}$ for these
binaries for the first time using optical, spatially-resolved photometry provided by HST.

\section{Brown Dwarf Binary Sample}\label{sample}

\subsection{Sample Selection}
Approximately 68 visual, ultracool binaries (spectral type M7 or later) in the solar
neighborhood have been identified in the last decade using high angular resolution imaging
surveys conducted with the Hubble Space Telescope (HST) and ground-based adaptive optics
systems \citep[e.g.,][]{Lane2002,Liu2008, Burgasser2007}.  Over half of these binaries have
been well-studied, undergoing extensive astrometric monitoring to determine precisely
measured total and individual masses (30-115 $M_{\mathrm{Jup}}$) and robust orbits have been
determined for 23 systems \citep{Konopacky2010,Dupuy2017}.  For a discussion on the larger
initial sample selection, see \citet{Burgasser2007} and for more information regarding
astrometric monitoring see \citet{Konopacky2010} and \citet{Dupuy2017}.

We selected seven binary systems from the set reported in \citet{Konopacky2010}. These
systems include 13 objects spanning spectral types from $\approx$L4-T5 and one late-M dwarf.  Since
the work published \citet{Konopacky2010}, distance and mass values have been updated for
many objects in our sample.  A summary of properties from the literature and updated masses
are provided in Table \ref{tab:bds}.  Distance uncertainties have been improved for all
objects in our sample.  Table \ref{tab:distance} shows previously used parallaxes with
updated distance calculations based on work from \citet{Dupuy2017} and the recent GAIA DR2
release \citep{Gaia2018}.

\begin{longrotatetable}
\begin{deluxetable*}{lllll llllll}
\tabletypesize{\footnotesize}
\tablecolumns{12}
\tablewidth{0pt}
\tablecaption{Properties of Binary Sample\label{tab:bds}}
\tablehead{
    &
    \multicolumn{4}{c}{Primary Component}&
    \multicolumn{4}{c}{Secondary Component}\\
    \cline{2-5} \cline{6-9}
    \colhead{System}&
    \colhead{$M_{\mathrm{Jup}}$}&
    \colhead{log(L$_{\mathrm{bol}}$/L$_{\bigodot}$)}&
    \colhead{SpT}&
    \colhead{$T_{\mathrm{eff}}$ (K)}&
    \colhead{$M_{\mathrm{Jup}}$}&
    \colhead{log(L$_{\mathrm{bol}}$/L$_{\bigodot}$)}&
    \colhead{SpT}&
    \colhead{$T_{\mathrm{eff}}$ (K)}&
    \colhead{Age (Gyr)}&
    \colhead{Ref.} 
    }
\startdata
HD 130948BC & 59.8$^{+2.0}_{-2.1}$ & -3.85 $\pm$ 0.06 & L4 $\pm$ 1 & 1920$^{+70}_{-60}$ & 55.6$^{+2.0}_{-1.9}$ & -3.96 $\pm$ 0.06 & L4 $\pm$ 1 & 1800$^{+50}_{-70}$ & 0.4-0.8 & 1,2,3\\
2MASS J0920+3517AB & 71 $\pm$ 5 & -4.270 $\pm$ 0.030 & L5.5 $\pm$ 1 & 1621 $^{+32}_{-30}$ & 116$^{+6}_{-8}$ & -4.340 $\pm$ 0.030 & L9 $\pm$ 1.5 & 1320 $\pm$ 250 & 3.1$^{+1.5}_{-1.7}$ & 1 \\
2MASS J1728+3948AB & 73 $\pm$ 7 & -4.29$^{+0.04}_{-0.05}$ & L5 $\pm$ 1 & 1600 $\pm$ 40 & 67 $\pm$ 5 & -4.49 $\pm$ 0.04 & L7 $\pm$ 1 & 1440 $\pm$ 40 & 3.4$^{+2.8}_{-2.1}$ & 1\\
2MASS J0850+1057AB$^{a}$ & 54 $\pm$ 8  & -4.22 $\pm$ 0.18 & L6.5 $\pm$ 1 &  1590 $\pm$ 290 & 54 $\pm$ 8 & -4.47 $\pm$ 0.18 & L8.5 $\pm$ 1.0 & 1380 $\pm$ 250 &  0.25-1.5 & 4 \\
LHS2397aAB & 93 $\pm$ 4 & -3.34 $\pm$ 0.04 & M8$\pm$ 0.5 & 2560 $\pm$ 50 & 66 $\pm$ 4 & -4.48 $\pm$ 0.04 & L7.5 $\pm$ 1 & 1440 $\pm$ 40 & 2.6$^{+0.6}_{-1.0}$ & 1\\
SDSS 1021-0304AB$^{b}$ & 52$^{+6}_{-7}$  & \nodata & T0 $\pm$ 1 & 1332 $\pm$ 100  & 52$^{+6}_{-7}$ & \nodata & T5 $\pm$ 0.5 & 1103 $\pm$ 100 & \nodata & 1,5 \\
2MASS J1534-2952AB & 51 $\pm$ 5 & -4.91 $\pm$ 0.07 & T4.5 $\pm$ 0.5 & 1150 $^{+40}_{-50}$ & 48 $\pm$ 5 & -4.99 $\pm$ 0.07 & T5 $\pm$ 0.5 & 1097 $\pm$ 50 & 3.0$^{+0.4}_{-0.5}$ &  1,2\\
\enddata
\tablecomments{\textbf{References} [1] \citet{Dupuy2017}, [2] \citet{Konopacky2010}, [3] \citet{Dupuy2014}, [4] \citet{Burgasser2010}, [5] \citet{Stephens2009} \\
$^{a}$ Total mass and temperature from \citet{Dupuy2017}. Luminosity reported from \citet{Konopacky2010}.
$^{b}$ Total mass from \citet{Dupuy2017}, and temperatures calculated from \citet{Stephens2009} spectral type-temperature relationship.}
\end{deluxetable*}
\end{longrotatetable}

\begin{deluxetable*}{lllll}
\tabletypesize{\footnotesize}
\tablecaption{Distances of Sample\label{tab:distance}}
\tablecolumns{5}
\tablewidth{0pt}
\tablehead{
    \colhead{Target}&
    \colhead{Distance [pc]}&
    \colhead{Parallax [mas]}&
    \colhead{Updated Distance [pc]}&
    \colhead{$\Delta$}
}
\startdata
HD 130948BC & 18.18 $\pm$ 0.08 & 55.73 $\pm$ 0.80 & 17.94 $\pm$ 0.25 & 1$\%$ \\ 
2MASS 0920+35AB & 24.3 $\pm$ 5.0 & 32.3 $\pm$ 0.6 & 31.0 $\pm$ 0.60 & 28$\%$ \\
2MASS 1728+39AB & 24.1 $\pm$ 2.1 & 36.4 $\pm$ 0.6 & 27.5 $\pm$ 0.47 & 14$\%$\\
2MASS 0850+10AB & 38.1 $\pm$ 7.3 & 31.4 $\pm$ 0.6 & 31.8 $\pm$ 0.55 & 17$\%$ \\
LHS 2397aAB & 14.3 $\pm$ 0.4 & 69.4903 $\pm$ 0.1760 & 14.3905 $\pm$ 0.0364 & 1$\%$\\ 
SDSS 1021-03AB & \nodata & 33.7 $\pm$ 1.2 & 29.7 $\pm$ 1.05 & \nodata\\
2MASS 1534-29AB & 13.59 $\pm$ 0.22 & 63.0 $\pm$ 1.1 & 15.9 $\pm$ 0.3 & 15$\%$
\enddata
\tablecomments{We provide updated distances using the most recent parallax values from
\citet{Gaia2018} and \citet{Dupuy2017}. A percent difference is calculated in column five
to show improvement in precision.  Note, SDSS 1021-03AB was not included in
\citet{Konopacky2010}.}
\end{deluxetable*}

\subsection{Photometry Measurements}\label{photometry}
The years of astrometric monitoring of the brown dwarf binary components in our sample resulted in precise, spatially resolved $J$, $H$, and $K$ broad-band flux measurements. The
majority of these near-infrared photometric measurements were obtained at the Keck
Observatory using the NIRC2 instrument behind adaptive optics. Details on the observing and
reduction procedure can be found in \citet{Konopacky2010}.

To obtain photometry and astrometry, we used the package StarFinder from
\citet{Diolaiti2000} (see Konopacky et al. 2010 for more details on the application for
this dataset). For photometry, StarFinder provides the ratio of the fluxes of the binary
components. We then use that ratio and the combined light magnitudes from 2MASS
\citep{Cutri2003} to derive individual apparent magnitude values for each component.
Uncertainties are calculated by computing the RMS of the photometry from all individual
images, and then propagating those uncertainties with those provided in the 2MASS catalog.

Additional photometric measurements were obtained from the Mikulski Archive for Space
Telescopes (MAST). Most of the archival data are from programs 10559/PI:Bouy and
9451/PI:Brandner using Advanced Camera for Surveys (ACS) and filters F625W, F775W, F850LP.
The archived data were supplemented with more recent observations using the Wide Field
Camera 3 (WFC3/UVIS), 11605/PI:Barman. The filters (F625W, F775W, and F850LP) were chosen
to provide Sloan-equivalent $r$, $i$, and $z$-band photometric measurements.

Data were collected in 2010 for Program 11605. We observed each target four times per
filter. Both raw and calibrated data frames were retrieved post observation from MAST. Data
are processed through the standard calibration pipeline for WFC3, as described in the WFC3
Data Handbook \citep{Gennaro2018}, including correction for geometric distortion.

Photometric measurements were obtained in two ways. First, we used the StarFinder algorithm
to derive positions and fluxes of the components. We ran this algorithm on the
distortion-corrected images. Since StarFinder requires a PSF estimate in the case of an
uncrowded image like that of a binary star, we used TinyTim to generate synthetic PSFs at
the proper wavelengths. Based on this PSF, StarFinder detected the components of the binary
and returned a centroid and estimated flux. Based on those fluxes, we used the WFC3/UVIS
zeropoints to compute the magnitudes of each source.

We also used the code written specifically for Hubble observations for photometry and
astrometry, $img2xym$ \citep{Anderson2006}, modified for WFC3/UVIS. It is designed to be
used with the ``FLT'' images, on which distortion has not been applied. It outputs the
positions and fluxes of stars that it identifies based on an isolation index which
describes the allowed separation between sources. Running this code on our frames provided
positions and fluxes, which were converted to magnitudes using the proper zeropoints.

Uncertainties in the WFC3/UVIS fluxes were determined as they were for NIRC2 data, by
fitting all data frames individually and then looking at the RMS variation. We found that
StarFinder and $img2xym$ returned consistent fluxes, and hence magnitudes, for most cases.
However, since a number of the binaries were very closely separated in the epoch of
observation (e.g., 2MASS 0920+35AB was separated by $<$1.5 pixels), we opted to present
here the results from the StarFinder analysis, as the $img2xym$ code warns about unreliable
results for stars separated by $<$2 pixels.

A complete list of photometry for our sample is provided in Tables \ref{tab:phot} and
\ref{tab:photIR}. In cases where the photometry was previously published, we have adopted
those values here.

\begin{deluxetable*}{lcccccccc}
\tabletypesize{\footnotesize}
\tablewidth{0pt}
\tablecaption{HST Photometric Observations\label{tab:phot}}
\tablehead{
    \multicolumn{1}{c}{Target}&
    \multicolumn{1}{c}{F625W}&
    \multicolumn{1}{c}{F625W}&
    \multicolumn{1}{c}{F775W}&
    \multicolumn{1}{c}{F775W}&
    \multicolumn{1}{c}{F814W}&
    \multicolumn{1}{c}{F850LP}&
    \multicolumn{1}{c}{F850LP}&
    \multicolumn{1}{c}{F1042W}\\
    \colhead{Instrument}&
    \colhead{[ACS]}&
    \colhead{[WFC3]}&
    \colhead{[ACS]}&
    \colhead{[WFC3]}&
    \colhead{[WFPC2]}&
    \colhead{[ACS]}&
    \colhead{[WFC3]}&
    \colhead{[WFPC2]}
}
\startdata
2MASS 0920+35A & \nodata & 21.14 $\pm$ 0.25 & \nodata & 18.35 $\pm$ 0.12 & 17.37 $\pm$ 0.18 & \nodata & 15.75 $\pm$ 0.18 & \nodata \\
2MASS 0920+35B & \nodata & 22.28 $\pm$ 0.71 & \nodata & 19.36 $\pm$ 0.36 & 18.25 $\pm$ 0.21 & \nodata & 16.07 $\pm$ 0.15 & \nodata\\
2MASS 1728+39A & \nodata & \nodata & \nodata & \nodata & 18.06 $\pm$ 0.07 & \nodata & \nodata & 15.60 $\pm$ 0.10  \\
2MASS 1728+39B & \nodata & \nodata & \nodata & \nodata & 18.71 $\pm$ 0.07 & \nodata & \nodata & 15.35 $\pm$ 0.12  \\
LHS 2397aA & \nodata & 17.66 $\pm$ 0.04 & \nodata & 15.75 $\pm$ 0.05 & 14.27 $\pm$ 0.04 & \nodata & 12.63 $\pm$ 0.05 & \nodata \\
LHS 2397aB & \nodata & 22.50 $\pm$ 0.33 & \nodata & 20.60 $\pm$ 0.34 & 18.69 $\pm$ 0.17 &\nodata & 16.78 $\pm$ 0.09 & \nodata \\
2MASS 0850+10A & 21.32 $\pm$ 0.28 & \nodata & 18.80 $\pm$ 0.24 & \nodata & 17.78 $\pm$ 0.15 & 16.56 $\pm$ 0.24 & \nodata & \nodata \\
2MASS 0850+10B & 22.62 $\pm$ 0.39 & \nodata & 19.96 $\pm$ 0.28 & \nodata & 19.25 $\pm$ 0.18 & 17.42 $\pm$ 0.26 & \nodata & \nodata  \\
SDSS 1021-03A & \nodata & \nodata & \nodata & 20.87 $\pm$ 0.63 & \nodata & \nodata & 17.01 $\pm$ 0.62 & \nodata \\
SDSS 1021-03B & \nodata & \nodata & \nodata & 22.03 $\pm$ 0.64 & \nodata & \nodata & 17.42 $\pm$ 0.62 & \nodata \\
2MASS 1534-29A & \nodata & \nodata & \nodata & 21.11 $\pm$ 0.14 & 19.22 $\pm$ 0.05 & \nodata & 16.64 $\pm$ 0.36 & 15.39 $\pm$ 0.12 \\
2MASS 1534-29B & \nodata & \nodata & \nodata & 21.61 $\pm$ 0.20 & 19.52 $\pm$ 0.06 & \nodata & 17.04 $\pm$ 0.15 & 15.59 $\pm$ 0.24
\enddata
\end{deluxetable*}

%
%
\begin{deluxetable}{lccc}
\tablecaption{Near-IR Photometric Observations\label{tab:photIR}}
\tablecolumns{3}
\tablewidth{0pt}
\tablehead{
    \colhead{Target}&
    \colhead{$J$}&
    \colhead{$H$}&
    \colhead{$K_{\mathrm{s}}$}
}
\startdata
HD 130948B & 12.53 $\pm$ 0.07 & 11.76 $\pm$ 0.10 & 10.98 $\pm$ 0.04\\
HD 130948C & 12.84 $\pm$ 0.08 & 12.05 $\pm$ 0.11 & 11.18 $\pm$ 0.04\\
2MASS 0920+35A & 13.89 $\pm$ 0.17 & 12.92 $\pm$ 0.07 & 12.12 $\pm$ 0.08 \\
2MASS 0920+35B & 13.94 $\pm$ 0.33 & 13.01 $\pm$ 0.09 & 12.44 $\pm$ 0.10 \\
2MASS 1728+39A & 14.27 $\pm$ 0.09 & 13.11 $\pm$ 0.08 & 12.20 $\pm$ 0.06 \\
2MASS 1728+39B & 14.58 $\pm$ 0.09 & 13.56 $\pm$ 0.08 & 12.80 $\pm$ 0.06 \\
LHS 2397aA & 11.18 $\pm$ 0.02 & 10.50 $\pm$ 0.03 & 10.02 $\pm$ 0.02\\
LHS 2397aB & 14.65 $\pm$ 0.08 & 13.61 $\pm$ 0.08 & 12.79 $\pm$ 0.04\\
2MASS 0850+10A & 14.24 $\pm$ 0.07 & 13.27 $\pm$ 0.11 & 12.56 $\pm$ 0.10 \\
2MASS 0850+10B & 14.70 $\pm$ 0.14 & 13.70 $\pm$ 0.14 & 12.91 $\pm$ 0.16 \\
SDSS 1021-03A & 14.22 $\pm$ 0.09 & 13.48 $\pm$ 0.09 & 13.27 $\pm$ 0.08\\
SDSS 1021-03B & 14.33 $\pm$ 0.09 & 14.27 $\pm$ 0.11 & 14.25 $\pm$ 0.08\\
2MASS 1534-29A & 14.57 $\pm$ 0.08 & 14.48 $\pm$ 0.07 & 14.46 $\pm$ 0.13\\
2MASS 1534-29B & 14.73 $\pm$ 0.12 & 14.83 $\pm$ 0.10 & 14.73 $\pm$ 0.15
\enddata
\end{deluxetable}

\begin{deluxetable}{llc}
\tabletypesize{\footnotesize}
\tablecaption{Spex Spectral Standards \label{tab:Standards}}
\tablecolumns{2}
\tablewidth{0pt}
\tablehead{
    \colhead{SpT}&
    \colhead{Template}&
    \colhead{References}
}
\startdata
M7 & VB 8 &  [1] \\ 
M8 & VB 10 &  [2] \\ 
M9 & LHS 2924 &  [3] \\ 
L0 & 2MASS J0345+2540 & [3] \\ 
L1 & 2MASSW J1439+1929 (opt), 2MASSW J2130-0845 &  [2,4] \\ 
L2 & 2MASSI J04082-1450 & [12] \\ 
L3 & 2MASSW J1146+2230 (opt), 2MASSW J1506+1321 &  [5,6] \\ 
L4 & 2MASS J21580457-1550098 &  [4] \\
L5 & DENIS-P J1228.2-1547 (opt), SDSS J083506.16+195304.4 & [5,7] \\ 
L6 & 2MASSs J0850+1057 (opt), 2MASSI J1010-0406 & [5,8]\\ 
L7 & DENIS-P J0205.4-1159 (opt), 2MASSI J0103+1935 & [5,9]  \\ 
L8 & 2MASSW J1632+1904 &  [6]   \\ 
L9 & DENIS-P J0255-4700 & [10] \\ 
T0 & SDSS J120747.17+024424.8 & [11] \\ 
T1 & SDSS J015141.69+124429.6 &  [2] \\ 
T2 & SDSSp J125453.90-012247.4 & [2] \\ 
T3 & 2MASS J1209-1004 & [2]\\ 
T4 & 2MASSI J2254188+312349 &  [2]  \\ 
T5 & 2MASS J15031961+2525196 &  [2] \\ 
T6 & SDSSp J162414.37+002915.6 &  [10] \\ 
T7 & 2MASSI J0727+1710 & [10]\\ 
T8 & 2MASSI J0415-0935 &  [2] \\ 
T9 & UGPS J072227.51-054031.2 &  [2] \\
\enddata
\tablecomments{A list of all brown dwarf templates used for spectral type determination
obtained from the Spex Library.  All templates had complete coverage from 0.8-2.5
$\micron$. Additional optical spectral standards were included for L1, L3, L5, L6, and L7
types due to their availability in the library (denoted as opt).\\
References: [1] \citet{Burgasser2008}, [2] \citet{Burgasser2004}, [3] \citet{BurgasserMcElwain2006}, [4] \citet{Kirkpatrick2010}, [5] \citet{Burgasser2010}, [6] \citet{Burgasser2007}, [7] \citet{Chiu2006}, [8] \citet{Reid2006}, [9] \citet{Cruz2004}, [10] \citet{Burgasser2006}, [11] \citet{Looper2008a}, [12] \citet{Cruz2017}
}
\end{deluxetable}

\section{Spectral Type Classification}\label{SPTClassification}
We derive spectral types for the nine objects in our sample using template fits to the SEDs
composed of resolved photometry for each object.  We then compare our inferred spectral
types to spectral types determined using spatially unresolved data and reported in the
literature \citep{Dupuy2012,Dupuy2017}.

We complied a set of brown dwarf template spectra listed in Table \ref{tab:Standards}
spanning spectral types M7 to T9 to compare to the objects in our sample.  Both optical and
near-IR spectral templates were used when available.  Template spectra come from the Spex
Prism Library collected by and maintained by \citet{Burgasser2014}.  Each spectral
template was absolute flux calibrated using 2MASS J-band photometry, and an absolute
magnitude was determined for each brown dwarf template for seven of our band-pass filters
for which there is complete spectral template coverage from 0.8-2.5 \microns.

\begin{equation}
M = -2.5 \ log\frac{\int f_{\lambda} \ S(\lambda) \ d\lambda}{\int f_{\mathrm{Vega}} \
S(\lambda) \ d\lambda},
\end{equation}

\noindent where $f_{\lambda}$ and $f_{\mathrm{Vega}}$ are the photon flux densities of the
template spectrum and Vega, respectively, at 10 parsecs.  $S(\lambda)$ is the filter
response function.

We used a $\chi^{2}$ approach to find the best match
\begin{equation}
\chi^{2} = \overset{n}{\underset{i}{\Sigma}} \frac{(M_{\mathrm{obs,i}}-M_{\mathrm{Spex,i}}+\delta M_{\mathrm{avg}})^{2}}{\sigma^{2}_{\mathrm{obs,i
}}},
\end{equation}
\noindent where $n$ is the number of photometric measurements available for a given object, M$_{\mathrm{obs}}$ are our observations, M$_{\mathrm{Spex}}$ are the Spex absolute magnitudes for each spectral template, $\delta$M$_{\mathrm{avg}}$ is the average offset between the observations and Spex magnitudes, and $\sigma_{\mathrm{obs}}$ are the uncertainties for our observations. The results of our fitting
and resulting component spectral types are provided in Table \ref{tab:SptFit}.

\begin{deluxetable*}{lcccc}
\tabletypesize{\footnotesize}
\tablecaption{Spectral Type Fitting \label{tab:SptFit}}
\label{tab:spexBF}
\tablewidth{0pt}
\tablehead{
    \colhead{Target}&
    \colhead{SpT Lit.}&
    \colhead{Fit SpT}&
    \colhead{Fit Name}&
    \colhead{$\chi^{2}$}
}
\startdata
HD 130948B & L4$\pm$1 & L3$\pm$1  & 2MASSW J1146345+223053 & 3.3 \\
HD 130948C & L4$\pm$1 & L5$\pm$1 & SDSS J083506.16+195304.4 & 5.3 \\
2MASS 0920+35A & L5.5$\pm$1 & L3$\pm$3 & SDSS J083506.16+195304.4 & 4.5\\
2MASS 0920+35B & L9 $\pm$1.5 & L4$\pm$3 & DENIS-P J0205.4-1159 & 0.2 \\
2MASS 1728+39A & L5 $\pm$1 & L7$\pm$1 & 2MASS J0103320+193536 & 5.4 \\
2MASS 1728+39B & L7 $\pm$1 & L9$\pm$1 & DENIS-P J0255-4700 & 5.4 \\
LHS 2397aA & M8$\pm$1 & M8$\pm$1 & VB 10 & 5.60\\
LHS 2397aB & L7.5$\pm$1 & L8$\pm$1 & 2MASSW J1632291+190441 & 2.1 \\
2MASS 0850+10A & L6.5$\pm$1 & L5$\pm$2 & SDSS J083506.16+195304.4 & 0.5\\
2MASS 0850+10B & L8.5$\pm$1 & L6$\pm$2  & 2MASSs J0850359+105716 & 3.1\\
SDSS 1021-03A & T0 $\pm$1 & T2$\pm$1 & SDSSp J125453.90-012247.4 & 5.2\\
SDSS 1021-03B & T5 $\pm$1 & T8$\pm$1 & 2MASSI J0415195-093506 & 1.6\\
2MASS 1534-29A & T4.5$\pm$1 & T5$\pm$1 & 2MASS J15031961+2525196 & 5.4\\
2MASS 1534-29B & T5$\pm$1 & T6$\pm$1 & SDSSp J162414.37+002915.6 & 6.6
\enddata
\tablecomments{Previously determined spectral types for objects in our sample are listed in
column two \citep{Dupuy2017}.  Columns 3-4 provide the best-fitting brown dwarf template
from Table \ref{tab:Standards}. HD 130948B and 2MASS 0850+10B were both fit to optical
templates for L3 and L6, respectively.}
\end{deluxetable*}

A lack of brown dwarf spectral templates with broad coverage across optical and
near-infrared wavelengths limited our ability to use all of our HST data. However, the
majority of spectral types we determined from our template fitting were consistent with
previously reported types from the literature.  Table \ref{tab:SptFit} shows a summary of
our results.  In a few cases for the coolest objects in our sample, we found a slightly
later classification by 1-2 spectral types than \citet{Dupuy2017}.

The 2MASS 0920+35AB system warrants further discussion.  2MASS 0920+35A was fit to an
earlier L3 $\pm$ 3 but was consistent with literature (L5.5 $\pm$ 1).  2MASS 0920+35B was
fit to L4 $\pm$ 3 discrepant from the literature value by 0.5 (L9 $\pm$ 1.5).  It was
difficult to determine a type for both components in this system due to similar $\chi^2$
values for L3, L4, L5, L7 types.  \citet{Dupuy2017} suggest 2MASS 0920+35B may be an
unresolved binary partially due to a large individual mass of $116^{+7}_{-8}$
M$_{\mathrm{Jup}}$.  The trends in spectral type and local minima further strengthen the
possibility that 2MASS 0920+35B is indeed composed of more than one object.

The consistency of our fitted spectral types to the types reported in the literature give
us confidence that our data, taken over multiple epochs and using multiple instruments, is
sound.  Given the large errors on some of our fitted spectral types; however, we will defer
to the more precise types reported by \citet{Dupuy2017} whenever a spectral type is
necessary.


\section{Model Atmospheres}
To investigate the properties of the brown dwarfs in our sample further, we use the model
atmosphere code $\PHOENIX$ to produce grids of synthetic spectra and photometry to compare
to our data.  We use the one-dimensional version of $\PHOENIX$ which self-consistently
calculates the atmospheric structure and emergent spectrum under the assumptions of
hydrostatic, chemical, and radiative-convective equilibrium \citep{Hauschildt1998}. 
$\PHOENIX$ solves the radiative transfer line-by-line and maintains a frequently updated
database of molecular opacities \citep{Tennyson2012}. The atmosphere models were calculated with a sampling of 1 \AA\ between 0.9 and 5 $\micron$. For the remaining wavelength ranges, the sampling varies from 2 to 5 \AA. For clarity, the model spectra have been convolved with a Gaussian of FWHM $\approx$ 70 \AA\ unless otherwise specified.

Solid and liquid particles suspended in an atmosphere (clouds) are arguably the most
complex process to include in brown dwarf and giant planet atmosphere models.  Atmospheres
of substellar-mass objects span a large range of temperatures and pressures that allow the
formation of a complex mixture of condensate species.  Early works tackling the issues of
cloud formation often focused on bracketing the extreme limits of condensate opacities by
exploring cloud-free and complete chemical equilibrium clouds.  Examples include the Cond
and Dusty models of \citet{Allard2001} and comparable models from \citet{Burrows2001}. 
These early works explained well the broad range of near-IR colors of brown dwarfs but were
not intended to reproduce the observed properties of all objects, especially not those in
or near the L/T transition.

A number of cloud models with higher degrees of parameterization were developed to expand
the applicability of atmosphere models for a greater variety of brown dwarfs suspected of
containing clouds.  Such models have included a range of tunable parameters for solid
particle growth, mixing, and sedimentation timescales as well as mean grain size, particle
size distributions, and explicit limits on position and vertical extent of the cloud (see
\cite{Marley2002} or \cite{Helling2008} for a comprehensive review).  For this work we are interested in quantifying the simplest set of cloud parameters capable of reproducing the SEDs of brown dwarfs across the L/T transition.  New atmosphere grids were developed specifically for
this study informed by previous exoplanet atmosphere grids
\citep{Barman2011,Barman2015,Stone2020,Miles2020} and after extensive testing of various
cloud properties.

\begin{figure}
\plotone{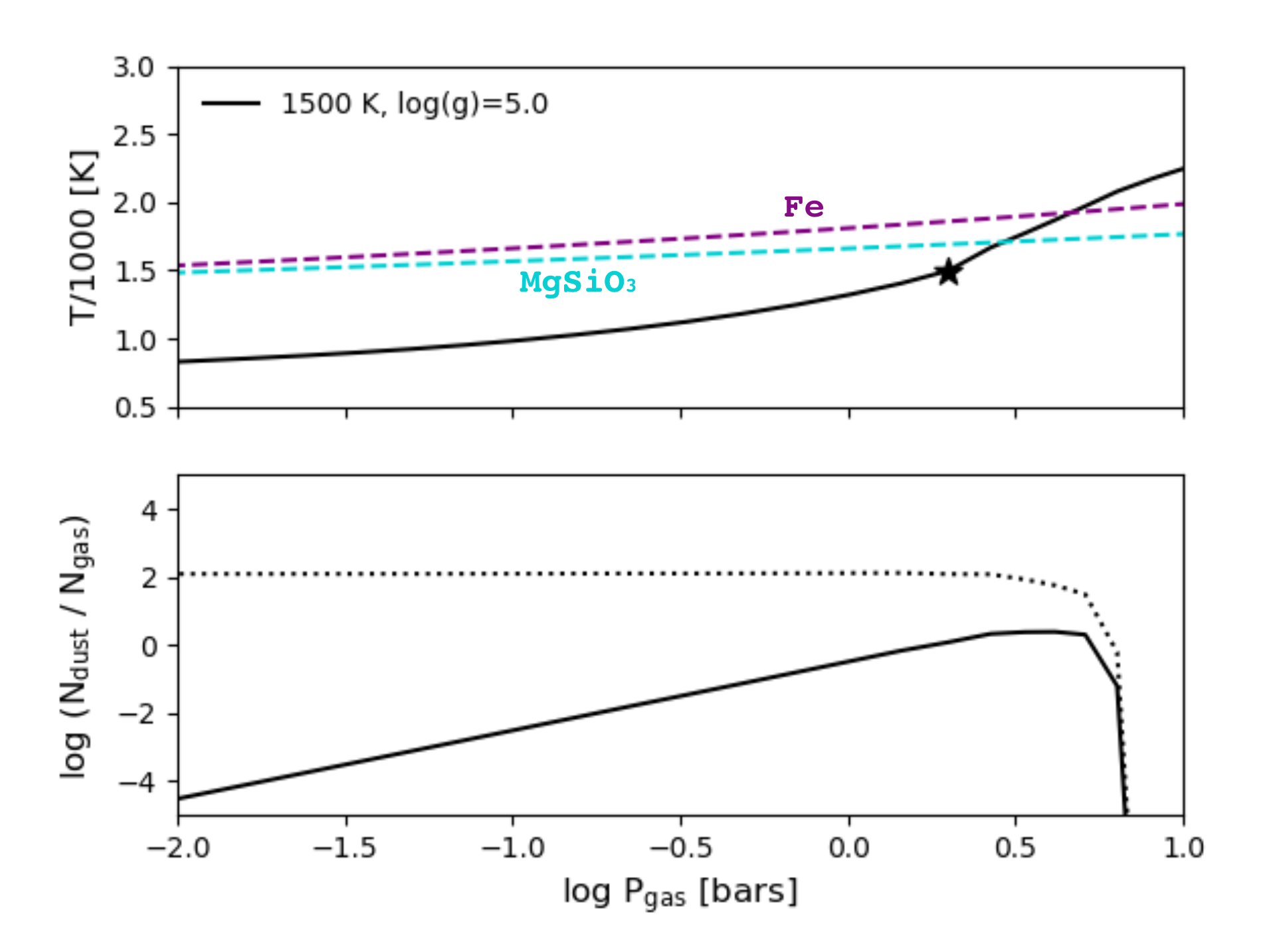}
\caption{Top: Temperature-Pressure profile for cloudy atmosphere model with P$_{c}$=20 bars and
a$_{0}$=0.25 \microns. Condensation curves for iron and enstatite are shown. The star represents the location of the photosphere. Bottom:
dust-to-gas ratio for pure equilibrium clouds (dotted line) and intermediate cloudy model
(solid line).\label{fig:TPClouds}}
\end{figure}

To that end we use the parameterized cloud model described in \citet{Barman2011}.  
The model includes one cloud composed of multiple condensates, each contributing to the total opacity based on their absorption and scattering cross-sections and relative number densities. A multiplicative weighting function is applied to the chemical equilibrium number densities where the value of the function is one for P$_{gas}$ $>$ P$_{c}$ and decreases exponentially for P$_{gas}$ $\leq$ P$_{c}$ (only a single P$_{c}$ value is used for each atmosphere model). This weighting function is similar to the family of models described in \citet{Burrows2006}; however, here the base of the cloud is always set by the deepest model layer where the chemical equilibrium condensate number density is non-zero. Figure \ref{fig:TPClouds} illustrates the basic structure for a model where P$_{c}$ is larger than P$_{gas}$ at the cloud base. In such a situation, the condensate number densities across the cloud layers are multiplied only by the exponentially decaying part of the weighting function and, thus, are less than the chemical equilibrium values across all cloud layers. For models where P$_{c}$ is smaller than P$_{gas}$ at the cloud base, the condensate number densities will be equal to their chemical equilibrium values for layers between P$_{c}$ and the cloud base, then dropping off for P$_{gas}$ $<$ P$_{c}$. With this simple model, we can adjust the cloud's vertical extent and particle number density with a single parameter.

\begin{figure}
\plotone{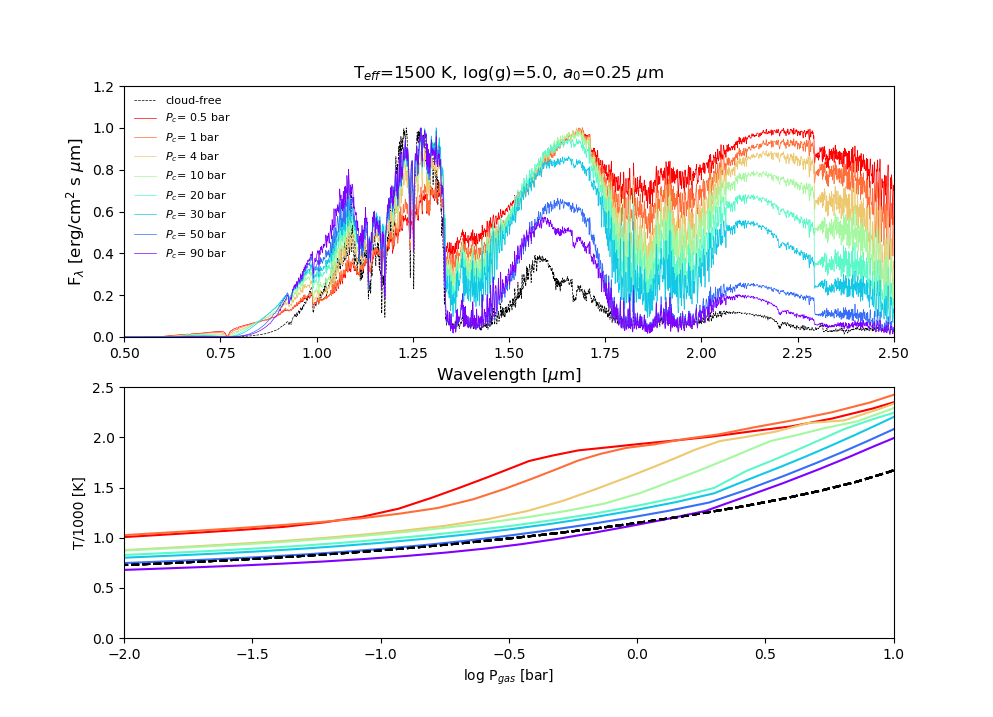}
\caption{Comparison between model spectra and T-P profiles when cloud pressure, P$_{c}$, is varied. A cloud-free model is provided in black. Spectra were normalized and convolved with a Gaussian of FWHM$\approx$12 \AA\ for clarity. All parameters are held constant except for cloud pressure. \label{fig:PGSPlot}}
\end{figure}

The cloud opacity is included self-consistently in the overall model calculation. Each model starts from a previous model calculation that is the closest to the new model across the various parameters. After each interation, the gas and condensate chemistry and associated opacities are recalculated based on a revised temperature structure. The process is repeated as the model converges toward radiative-convective equilibrium. Figure \ref{fig:PGSPlot} illustrates how cloud pressure influences the spectrum and temperature-pressure profile.

\begin{figure}
\plotone{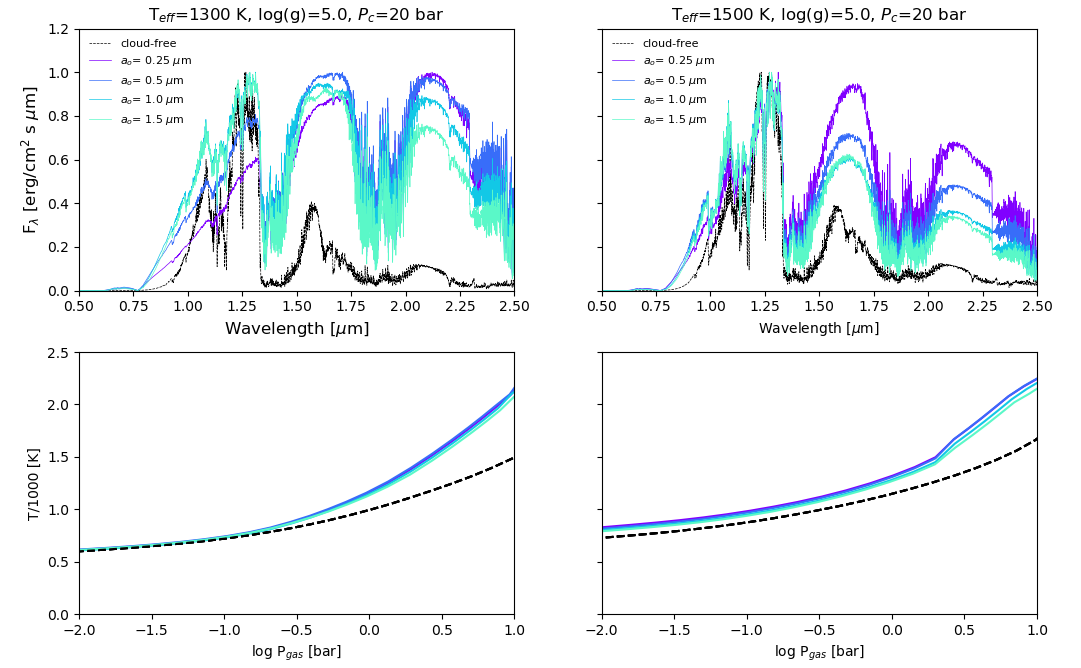}
\caption{Comparison between model spectra and T-P profiles when mean grain size is varied. A cloud-free model is provided in black. Normalized spectra are shown for two effective temperatures. Spectra are convolved with a Gaussian of FWHM$\approx$12 \AA\ for clarity. All parameters held constant except for
grain size in each panel.\label{fig:GSPlot}}
\end{figure}

For simplicity, grain size is assumed to be independent of height and governed by an
additional free parameter of mean particle size, $a_{0}$.  A log-normal distribution of
grain sizes is included based on work from \citet{Marley1999}. The range of mean particle
sizes in our models span 0.25-10 $\micron$.  Model grains include those in the PHOENIX
database that are thermodynamically permissible and included in the total cloud opacity
\citep{Ferguson2005}.  Figure \ref{fig:GSPlot} illustrates how grain size influences the
spectrum when P$_{c}$ is held constant at 20 bars.  At 1500 K adjustments to grain size
have little impact at the $J$ band near 1.3 \microns.  However, as objects approach the L/T
transition and the temperature is decreased to 1300 K, changes in $J$-band flux are more
apparent and large differences in spectral shape can be seen between the $H$ and $K$ bands
when comparing cloudy versus cloud-free models.

\section{Results} \label{sec:results}
\subsection{Grid Fitting} \label{GridFitting}
We fit resolved photometry to model grids to better understand the nature of clouds in
substellar atmospheres spanning the L/T transition.  Table \ref{tab:ModelParams}
summarizes the grid properties and includes the new grids created specifically for this
work, extending cloud properties across a broader range of pressures and incorporating
models with smaller mean grain sizes.  All models in the grid are solar metallicity with
effective temperatures ranging from 800-2000 K and surface gravities of log(g)=3.0-5.5. 
The conservative range of our models is suitable for objects from early L through early T
dwarfs \citep{Kirkpatrick2005}.

We calculate synthetic photometry for all models using HST and 2MASS filter transmission
curves mentioned in Section \ref{sample}.  A best-fitting model is determined for each
object using a $\chi^{2}$ fitting approach.  We take the scale factor as
\begin{equation}
s = (R/D)^2,
\end{equation}

\noindent where R is the radius of the object to be fit, and D is 10 parsecs for absolute
magnitude.


\begin{deluxetable*}{lllll}
\tabletypesize{\footnotesize}
\tablecaption{Model Parameter Ranges \label{tab:ModelParams}}
\tablewidth{0pt}
\tablehead{
    \multicolumn{1}{c}{$T_{\mathrm{eff}}$}&
    \multicolumn{1}{c}{log(g)}&
    \multicolumn{1}{c}{P$_{\mathrm{c}}$}&
    \multicolumn{1}{c}{a$_{0}$}&
    \multicolumn{1}{c}{Increments}\\
    \colhead{[K]}&
    \colhead{[cm $s^{-2}$]}&
    \colhead{[bar]}&
    \colhead{[$\mu$m]}&
    \colhead{}
 }
\startdata
800-2000 & 3.5-5.5 & 0.5, 1, 4 & 1 & 100 K; 0.5 in log(g) \\
800-2000 & 4.75, 5.0, 5.5 & 10, 20, 30 & 0.25, 0.5, 1 & 100 K \\
1700-2000 & 4.5-5.5 & 0.1, 0.5 & 0.25 & 100 K; 0.5 in log(g)
\enddata
\end{deluxetable*}

Table \ref{tab:Fits} shows the results of our model fitting.  Models are allowed at the
68\% confidence level by taking $\chi^{2}$ distributed with five degrees of freedom (the
four atmosphere model parameters in Table \ref{tab:ModelParams} and the object's radius,
which is simultaneously fit).  This includes all models with $\Delta \chi^{2}<$ 11.3. 
Listed first is the overall best-fitting model from the cloudy atmosphere grids.  We then
calculate the weighted mean parameters by
\begin{equation}
\bar{m} = \frac{\Sigma_i W_i m_i}{\Sigma_i W_i},
\end{equation}

\noindent where the weight $W_{i}$ for each model $m_{i}$ is given by
\begin{equation}
W_i = H_{i} e^{-0.5\chi^2},
\end{equation}

\noindent where $H_{i}$ is a scaling factor that accounts for the uneven spacing of some
atmospheric parameters on the grid, down weighting regions of dense sampling and
up weighting regions of coarse sampling in proportion to the amount of parameter space
covered.

We used the same approach as \citet{Stone2016} to calculate uncertainties using sided
variance estimates with
\begin{equation}
\sigma_{m} = \frac{\Sigma_{i} W_i (m_i - \bar{m})^2}{\Sigma_{i} W_i},
\end{equation}

\noindent where the sum is calculated using parameters above (+) or below (-) the mean
values.  In cases where the edge of the grid boundary was approached, we report an
upper/lower limit for our uncertainty.


\begin{deluxetable*}{llllll lllll}
\tabletypesize{\footnotesize}
\tablecaption{Model Atmosphere Fits \label{tab:Fits}}
\tablecolumns{11}
\tablehead{
    &
    \multicolumn{5}{c}{Best Fit}&
    \multicolumn{5}{c}{Weighted Mean}\\
    \cline{2-6} \cline{7-11} 
    \colhead{Name}&
    \colhead{T$_{\mathrm{eff}}$}&
    \colhead{log(g)}&
    \colhead{R$_{\mathrm{Jup}}$}&
    \colhead{P$_{\mathrm{c}}$}&
    \colhead{a$_{0}$}&
    \colhead{T$_{\mathrm{eff}}$}&
    \colhead{log(g)}&
    \colhead{R$_{\mathrm{Jup}}$}&
    \colhead{P$_{\mathrm{c}}$}&
    \colhead{a$_{0}$}
}
\startdata
HD 130948B & 1400 & 3.5 & 1.68 & 4 & 1.0 & 1903$^{+97}_{-4}$ & 4.5$^{+0.50}_{-1.0}$ & 1.04$^{+0.27}_{-0.29}$ & 0.22$^{+0.28}_{-0.12}$ & $\leq$ 1.0 \\
HD 130948C & 1400 & 4.0 & 1.76 & 4 & 1.0 & 1808$^{+92}_{-8}$ & 4.5$^{+0.59}_{-1.0}$ & 1.03$^{+0.16}_{-0.34}$ & 0.47$^{+0.03}_{-0.67}$ & $\leq$ 1.0 \\
2MASS 0920+35A & 1300 & 4.75 & 1.50 & 4 & 1.0 & 1517$^{+331}_{-194}$ & 4.88$^{+0.47}_{-0.72}$ & 1.41$^{+0.34}_{-0.47}$ & 2.9$^{+2.9}_{-2.6}$ & 0.70$^{+0.30}_{-0.43}$\\
2MASS 1728+39A & 1500 & 5.5 & 0.95 & 30 & 0.25 & 1487$^{+29}_{-131}$ & 5.17$^{+0.33}_{-0.32}$ & 0.97$^{+0.13}_{-0.05}$ & 20$^{+10}_{-9}$ & 0.26$^{+0.68}_{-0.01}$  \\
2MASS 1728+39B & 1400 & 4.75 & 0.83 & 20 & 1.0 & 1420$^{+100}_{-22}$ & 4.99$^{+0.43}_{-0.83}$ & 0.89$^{+0.06}_{-0.09}$ & 8$^{+10}_{-5}$ & 0.93$^{+0.07}_{-0.44}$ \\
LHS2397aB & 1300 & 4.75 & 0.95 & 30 & 1.0 & 1413$^{+93}_{-102}$ & 4.84$^{+0.23}_{-0.09}$ & 0.81$^{+0.12}_{-0.11}$ & 23$^{+7}_{-3}$ & 0.54$^{+0.46}_{-0.27}$  \\
2MASS 0850+10A & 1800 & 5.5 & 0.56 & 10 & 0.25 & 1717$^{+91}_{-207}$ & 5.08$^{+0.42}_{-0.63}$ & 0.65$^{+0.22}_{-0.10}$ & 7.2$^{+4.2}_{-6.1}$ & 0.36$^{+0.55}_{-0.11}$  \\
2MASS 0850+10B & 1500 & 5.5 & 0.74 & 20 & 0.25 & 1423$^{+80}_{-277}$ & 5.3$^{+0.20}_{-0.69}$ & 0.93$^{+0.80}_{-0.18}$ & 14$^{+10}_{-7}$ & 0.37$^{+0.63}_{-0.12}$ \\
SDSS 1021-03A & 1700 & 5.0 & 0.50 & 30 & 0.25 & 1668$^{+37}_{-97}$ & 4.92$^{+0.09}_{-0.17}$ & 0.52$^{+0.09}_{-0.03}$ & 28$^{+20}_{-8}$ & 0.36$^{+0.57}_{-0.11}$\\
SDSS 1021-03B & 1700 & 4.75 & 0.36 & 30 & 1.0 & 1698$^{+2}_{-98}$ & $\leq$ 5.5 & 0.36$^{+0.06}_{-0.02}$ & $\geq$ 0.1 & 0.96$^{+0.04}_{-0.57}$\\
2MASS 1534-29A & 1700 & 5.5 & 0.38 & 30 & 0.25 & 1699$^{+1}_{-99}$ & $\geq$ 3.5 & 0.38$^{+0.02}_{-0.01}$ & 24$^{+7}_{-4}$ & $\leq$ 1.0\\
2MASS 1534-29B & 1700 & 5.5 & 0.32 & 20 & 0.25 & 1699$^{+1}_{-99}$ & $\geq$ 3.5 & 0.33 $\pm$ 0.01 & 21$^{+9}_{-3}$ & $\leq$ 1.0
\enddata
\tablecomments{Best-fitting grid models and weighted means with 2-$\sigma$ uncertainties
are provided. Weighted mean fits for HD 130948BC use a fixed range for radius 0.8-1.2
R$_{\mathrm{Jup}}$ constrained by \citet{Saumon2008} evolutionary models.  Fits in the table for
2MASS 1728+39B exclude the F814W photometric point and are discussed further in Section
\ref{sec:1728}.}
\end{deluxetable*}

\begin{figure*}
\gridline{\fig{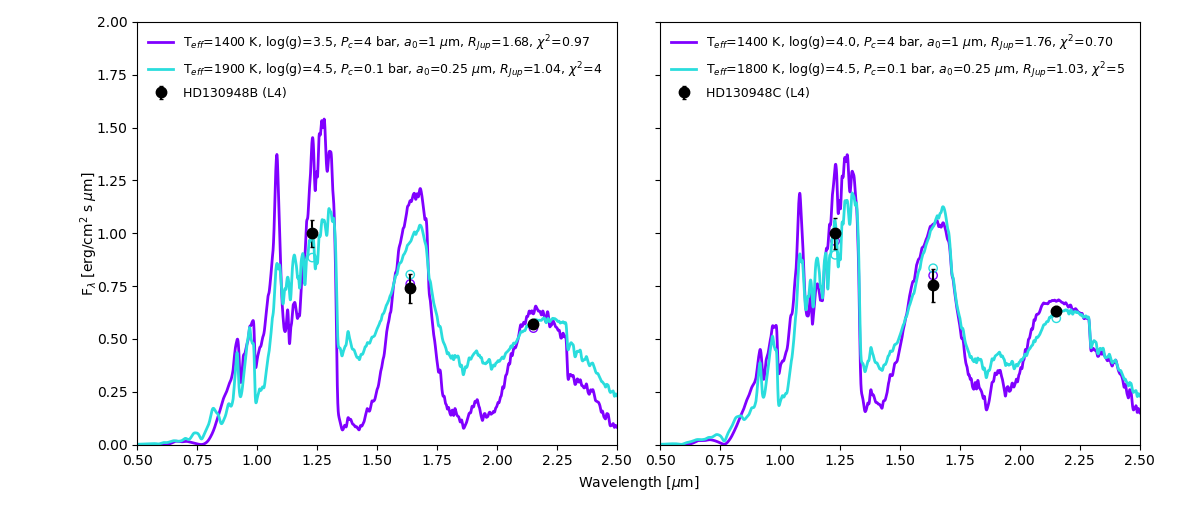}{0.80\textwidth}{(a)}}
\gridline{\fig{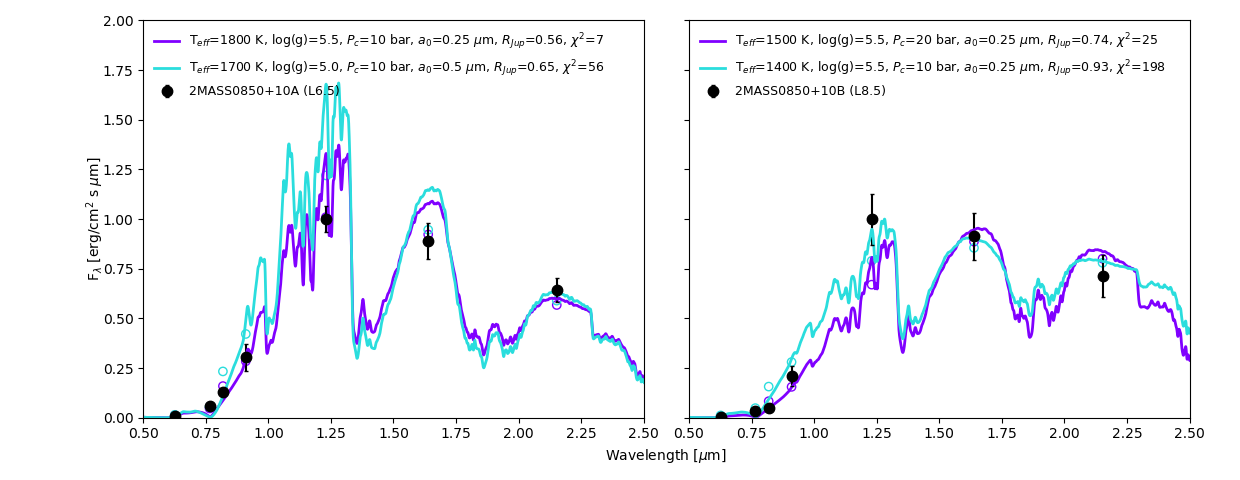}{0.80\textwidth}{(b)}}
\gridline{\fig{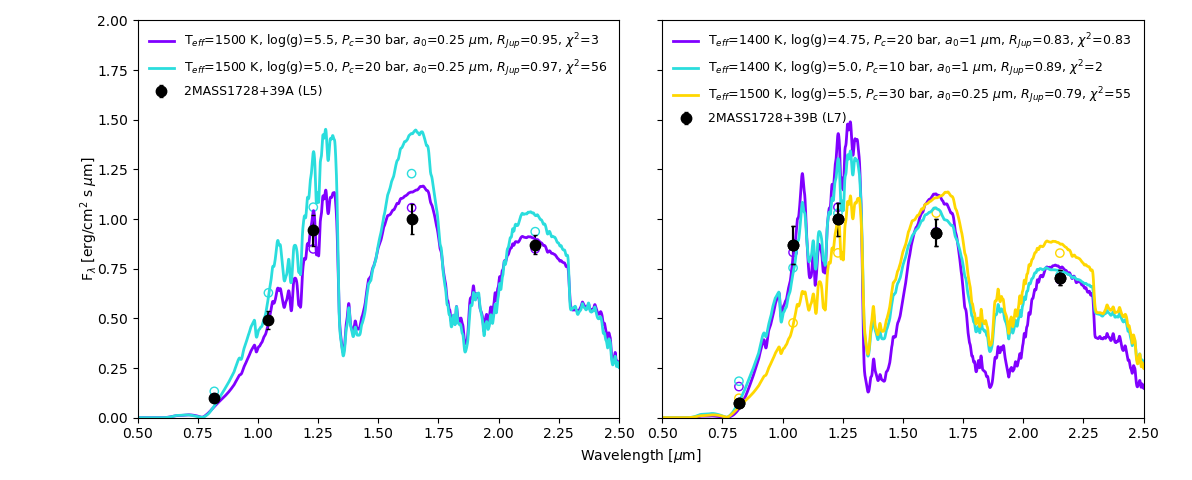}{0.80\textwidth}{(c)}}
\caption{Best-fit (purple) and weighted mean (cyan) atmosphere models from Table
\ref{tab:Fits}. We chose the closest available model from our grid to the mean parameters.
Corresponding synthetic photometry is plotted in same color as model spectra. Surface
gravity is in cm $s^{-2}$. Spectral types included are from \citet{Dupuy2017}.
\label{fig:BFPlots}}
\end{figure*}
\begin{figure*}
\gridline{\fig{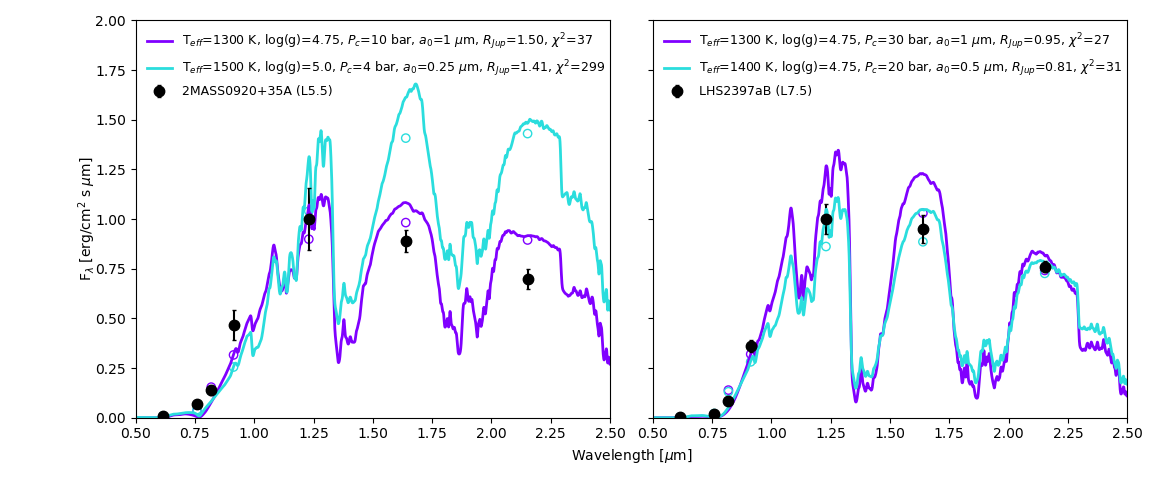}{0.85\textwidth}{(a)}}
\gridline{\fig{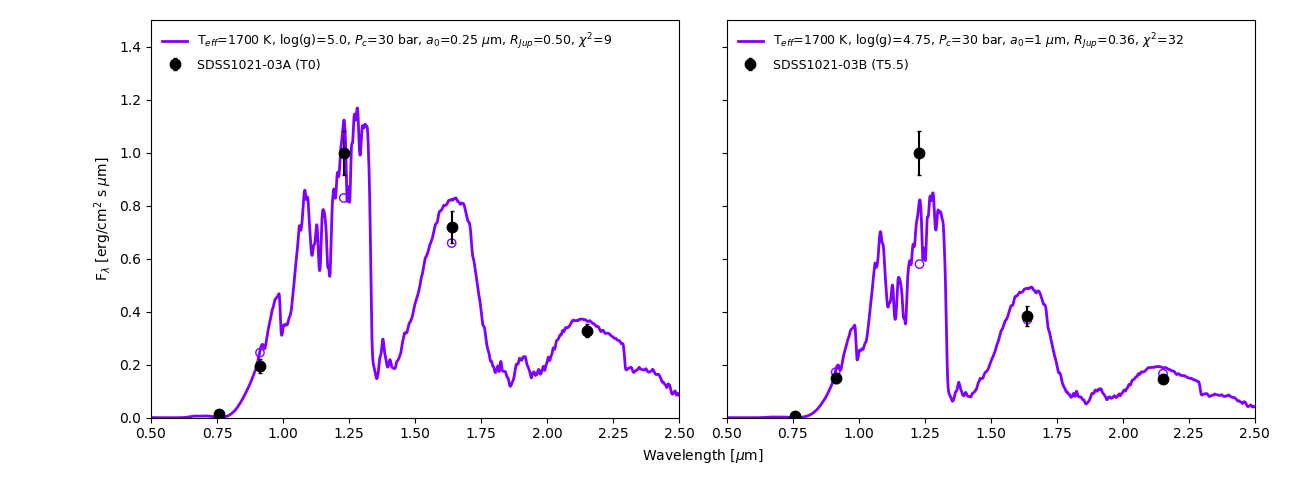}{0.85\textwidth}{(b)}}
\gridline{\fig{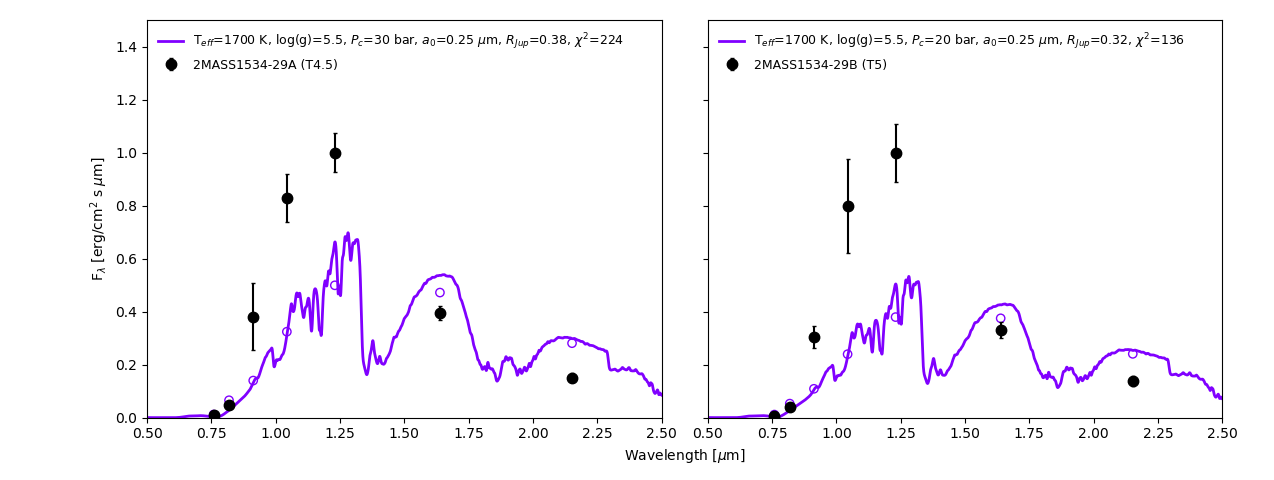}{0.85\textwidth}{(c)}}
\caption{Best-fit (purple) and weighted mean (cyan) atmosphere models from Table
\ref{tab:Fits}. Only the best fits are shown for SDSS 1021-03AB and 2MASS 1534-29AB
because weighted mean parameters are similar. Surface
gravity is in cm $s^{-2}$. Spectral types included are from \citet{Dupuy2017} \label{fig:BFPlots2}}
\end{figure*}

The results of our analysis are shown in Figures \ref{fig:BFPlots} and \ref{fig:BFPlots2}
for our sample of objects.  We plot the best-fit and weighted mean grid models for each
object from Table \ref{tab:Fits}.  These fits highlight the role clouds play in the
uncertainties of effective temperature, gravity, and radius when fitting cloudy model
grids to photometric observations.  Three objects, 2MASS 1728+39AB and LHS 2397aB, were
fit well by the grid and had properties consistent with evolutionary models shown in
Figures \ref{fig:BFPlots} (c) and \ref{fig:BFPlots2} (a).  

Our fit of five parameters to HD 130948BC is not adequately constrained by the three
photometric points used here.  To help guide our fitting in subsequent sections and
approach a unique solution for these objects, we applied a prior on radius (0.8-1.2
R$_{\mathrm{Jup}}$) from evolutionary models for the weighted mean fits in Table \ref{tab:Fits}.
The unconstrained best-fit model and weighted fit are provided in Figure \ref{fig:BFPlots}
(a).  This shows how two atmosphere models with very different effective temperatures,
gravities, and cloud properties can match $JHK$ photometric observations well.  The best
fits plotted in purple are much cooler than expected for early L-type objects and approach
non-physical radii ($\geq$ 1.5 $R_{\mathrm{Jup}}$).

We note a few other objects where the grid fits present tension: 2MASS 0850+10AB, 2MASS
0920+35A, SDSS 1021-03AB, 2MASS 1534-29AB.  Without incorporating evolutionary constraints
on radius for the aforementioned systems,  models allowed at 2-$\sigma$ over or
under-predict effective temperature expected for a given spectral type \citep{Leggett2002,
Nakajima2004} and can result in non-physical radii. We believe this may be influenced by an unresolved component for some objects (e.g., 2MASS 0850+10A), and the cloud parameter space used in our grids does not appear to be appropriate for early-to-mid T dwarfs (e.g., 2MASS 1534-29AB). Figure \ref{fig:BFPlots} shows the grid fits for 2MASS 0850+10AB,
SDSS 1021-03AB, and 2MASS 1534-29AB are warmer ($T_{\mathrm{eff}}$ $\geq$ 1700 K) and
approach smaller than expected radii ($\leq$ 0.70 $R_{\mathrm{Jup}}$) for objects of field
age \citep{Dupuy2017}.  Conversely, the model fits for 2MASS 0920+35A shown in Figure
\ref{fig:BFPlots2} have radii larger than expected ($\geq$ 1.4 $R_{\mathrm{Jup}}$) for
mid-L dwarfs \citep{Leggett2002, Nakajima2004}.

\section{Evolutionary Model Comparisons}\label{EvoComparison}
The addition of cloud properties in atmosphere models can lead to further degeneracies in
mass and radius resulting in good fits to photometric data but producing non-physical
properties given what we know of brown dwarf physics.  An effective way to evaluate the
quality of atmosphere model fits is by comparing the results to the bulk properties from
evolutionary model predictions.  We use bolometric luminosity constrained from our grid
fits combined with the measured masses \citep{Dupuy2017, Konopacky2010} to derive
evolutionary predictions. We derive evolutionary predictions using \citet{Saumon2008}
hybrid grids for mid-to-late L dwarfs in our sample and use the Cond grids from
\citet{Baraffe2003} for T dwarfs with T$_{\mathrm{eff}}$ $\lesssim$ 1300 K.  Evolutionary
properties are given in Table \ref{tab:Evo} and individual systems are discussed below.

We take comparisons a step further by running additional atmosphere models using a fixed
value for \textbf{effective} temperature, gravity, and radius from evolutionary predictions in Table
\ref{tab:Evo}.  Holding evolutionary properties constant, we ran additional models around
the best-fitting cloud properties determined from our grid fits in Section
\ref{sec:results}. The goal was to determine if our cloudy atmosphere models could fit
data well and remain consistent with substellar evolutionary model predictions.

Because we know the masses for these binary systems, distinguishing between different
atmospheric fits becomes more reliable.  We are able to eliminate model fits that may
represent the data well yet result in implied masses that deviate from empirical
observations.  Implied mass is calculated from surface gravity and radius for a given
atmosphere model.  All model fits for individual systems are discussed below.

\begin{longrotatetable}
\begin{deluxetable*}{llllll llllll}
\tabletypesize{\footnotesize}
\tablecaption{Evolutionary Model Predictions \label{tab:Evo}}
\tablecolumns{12}
\tablehead{
    &
    \multicolumn{5}{c}{Primary}&
    \multicolumn{5}{c}{Secondary}\\
    \cline{2-6} \cline{7-11}
    \colhead{System}&
    \colhead{M$_{\mathrm{Jup}}$}&
    \colhead{log($L_{\mathrm{bol}}/L_{\odot}$)}&
    \colhead{T$_{\mathrm{eff}}$}&
    \colhead{R$_{\mathrm{Jup}}$}&
    \colhead{log(g)}&
    \colhead{M$_{\mathrm{Jup}}$}&
    \colhead{log(${L_{\mathrm{bol}}}/{L_{\odot}}$)}&
    \colhead{T$_{\mathrm{eff}}$}&
    \colhead{R$_{\mathrm{Jup}}$}&
    \colhead{log(g)}&
    \colhead{Age(Gyr)}
}
\startdata
HD 130948BC & 59$\pm$1 & -3.85 $\pm$ 0.09 & 1916$^{+94}_{-90}$ & 1.05 $\pm$ 0.02 & 5.12 $\pm$ 0.02 & 56$^{+2}_{-1}$ & -3.91 $\pm$ 0.05 & 1851$^{+46}_{-45}$ & 1.05 $\pm$ 0.02 & 5.10$^{+0.02}_{-0.03}$ & 0.42$^{+0.06}_{-0.04}$ \\
2MASS 0920+35AB & 71$^{+3}_{-4}$ & -4.28 $\pm$ 0.05 & 1604$^{+51}_{-52}$ & 0.91$^{+0.03}_{-0.01}$ & 5.32$^{+0.03}_{-0.04}$ & 58$^{+9}_{-7}$ & -4.70 $\pm$ 0.05 & 1260$^{+57}_{-52}$ & 0.91 $\pm$ 0.04 & 5.24 $\pm$ 0.10 & 1.82$^{+0.35}_{-0.18}$ \\
2MASS 1728+39AB & 71$^{+1}_{-2}$ & -4.37 $\pm$ 0.04 & 1536$^{+30}_{-41}$ & 0.90$^{+0.02}_{-0.01}$ & 5.34$^{+0.01}_{-0.04}$ & 68$^{+1}_{-4}$ & -4.52 $\pm$ 0.06 & 1417$^{+42}_{-58}$ & 0.89$^{+0.03}_{-0.01}$ & 5.33$^{+0.01}_{-0.05}$ & 2.82$^{+0.42}_{-1.24}$ \\
LHS 2397aAB & 92$\pm$1 & -3.34$\pm$0.04 & 2533$^{+41}_{-40}$ & 1.07 $\pm$ 0.02 & 5.29 $\pm$ 0.01 & 67$^{+1}_{-9}$ & -4.62 $\pm$ 0.02 & 1342$^{+10}_{-39}$ & 0.88$^{+0.04}_{-0.01}$ & 5.33$^{+0.01}_{-0.10}$ & 3.02$^{+0.07}_{-1.47}$ \\
2MASS 0850+10AB & 28$^{+4}_{-5}$ & -4.49 $\pm$ 0.06 & 1272$^{+58}_{-57}$ & 1.14$^{+0.03}_{-0.05}$ & 4.72$^{+0.11}_{-0.12}$ & 26$^{+5}_{-4}$ & -4.54 $\pm$ 0.09 & 1234$^{+79}_{-77}$ & 1.15$^{+0.03}_{-0.06}$ & 4.70$^{+0.09}_{-0.11}$ & 0.25$^{+0.13}_{-0.09}$ \\
SDSS 1021-03AB & 29 $\pm$ 4 & -4.70 $\pm$ 0.05 & 1219$^{+48}_{-47}$ & 1.00$^{+0.04}_{-0.03}$ & 4.85$^{+0.09}_{-0.08}$ & 23 $\pm$ 4 & -4.99 $\pm$ 0.01 & 1023$^{+20}_{-24}$ & 1.02$^{+0.04}_{-0.03}$ & 4.74$^{+0.08}_{-0.11}$ & 0.47$^{+0.08}_{-0.15}$\\
2MASS 1534-29AB & 52 $\pm$ 3 & -4.94 $\pm$ 0.02 & 1157$^{+21}_{-7}$ & 0.83$^{+0.02}_{-0.01}$ & 5.27$^{+0.03}_{-0.04}$ & 47 $\pm$ 3 & -5.08 $\pm$ 0.02 & 1062$^{+19}_{-9}$ & 0.85$^{+0.01}_{-0.02}$ & 5.21 $\pm$ 0.04 & 2.40 $\pm$ 0.07
\enddata
\tablecomments{Evolutionary model-derived properties from \citet{Saumon2008} hybrid grids.
For the T dwarfs, SDSS 1021A+B and 2MASS 1534-29A+B, evolutionary properties are from
\citet{Baraffe2003} COND grids. Luminosity for LHS 2397aA is from \citet{Dupuy2017}.}
\end{deluxetable*}
\end{longrotatetable}

\subsection{HD 130948B+C}
The best grid fit for the HD 130948B+C system does not match evolutionary-derived
properties.  This is not surprising because previous work has shown discrepancies between
atmosphere grids and evolutionary properties ($\pm$ 250 K) when only $J$, $H$, and $K$
photometry are used \citep{Dupuy2009a}.  \citet{Barman2011} and other groups find similar
issues for directly imaged planets.  Additionally, limited SED coverage for this system
leads to several well-fit models given grid fits are formally under-constrained.  However,
by fixing effective temperature, gravity, and radius using evolutionary-derived values we
can fit for two cloud parameters to arrive at a more reliable solution that is formally
allowed.

\begin{figure*}
\plotone{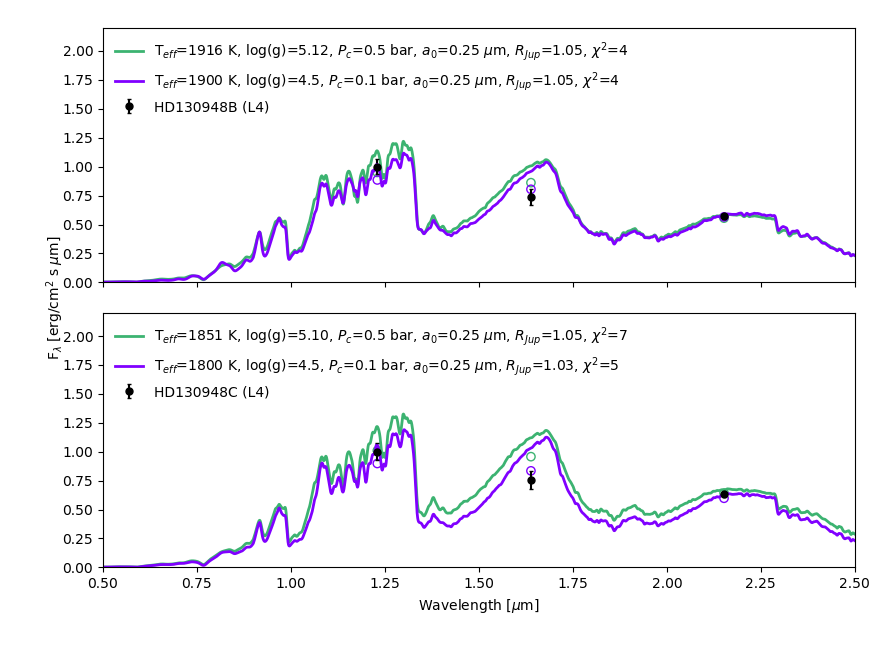}
\caption{New evolutionary fit atmosphere models (green) compared to weighted mean grid
fits (purple) for HD 130948B (top) and HD 130948C (bottom). Surface gravity is in cm
$s^{-2}$. \label{fig:HD130948BC}}
\end{figure*}

In Figure \ref{fig:HD130948BC} we show our previous weighted mean grid fit compared to our
new evolutionary fit with fixed $T_{\mathrm{eff}}$, log(g), and $R_{\mathrm{Jup}}$.  A
good fit to the data can be achieved for HD 130948B with a higher cloud and $a_{0}$=0.25
$\micron$ grain size.  The fits are nearly identical, but the grid fit with lower surface
gravity is inconsistent with measured mass compared to the higher surface gravity model
($M_{\mathrm{Jup}}$=14.08 and 58.68, respectively).  HD 130948C is slightly cooler than
the B component yet fit well with the same type of cloud and grain size.  Again, the lower
surface gravity model can be ruled out by implied mass ($M_{\mathrm{Jup}}$=13.54).  The
new evolutionary fits presented here for both objects are consistent with recent
atmospheric properties determined by \citet{Briesemeister2019} which included both $J$,
$H$, and $K$ photometry with additional ALES L-band spectra from 2.9–4.1 \microns.  Our
updated fits provide additional constraints on the cloud properties of early-to-mid L type
dwarfs.

\subsection{2MASS 0850+10AB}
\begin{figure*}
\plotone{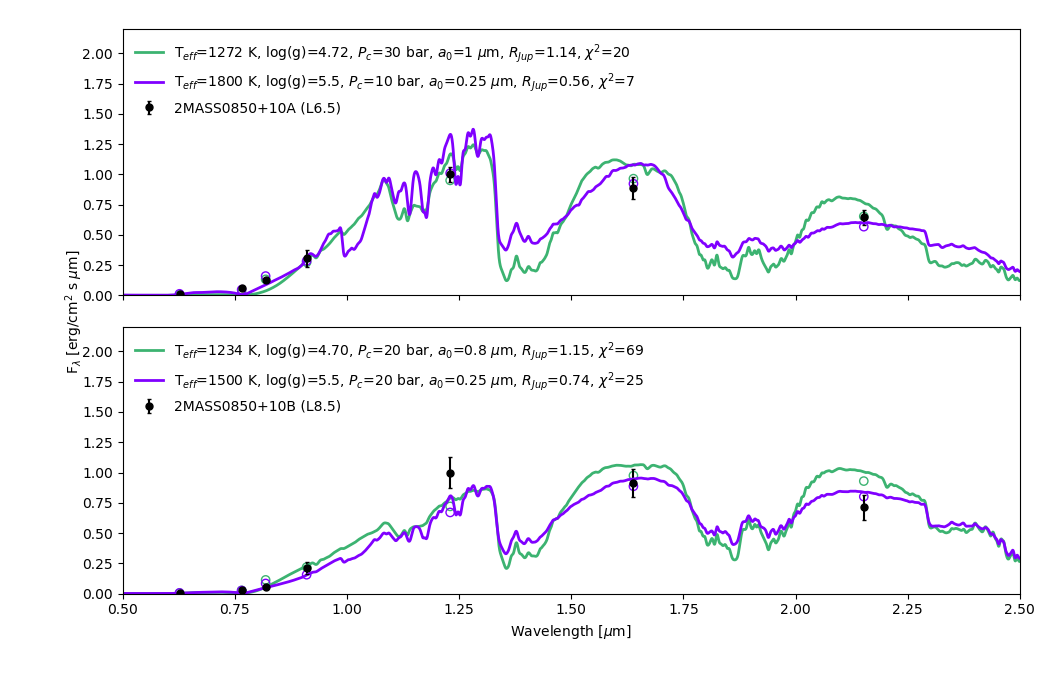}
\caption{New evolutionary fit atmosphere models (green) compared to best grid fits
(purple) for 2MASS 0850+10A (top) and 2MASS 0850+10B (bottom). Surface gravity is in cm
$s^{-2}$. \label{fig:2M0850AB}}
\end{figure*}

We show our new evolutionary fits compared to our previous best-fitting grid model in
Figure \ref{fig:2M0850AB} for 2MASS 0850+10A and B.  The initial grid fits significantly
over-predicted temperature ($\approx$ 250-500 K) and gravity ($\approx$ 0.75 dex) while
under-predicting radius compared to evolutionary predictions. Weighted mean parameters were consistent with evolutionary predictions for the secondary component but not for the primary (Table \ref{tab:Fits}). Table \ref{tab:Evo} shows evolutionary predictions for this system have the lowest values of gravity (log(g) $\leq$ 4.75) for all the L dwarfs studied in this sample.

2MASS 0850+10A is fit well when bulk properties are fixed to evolutionary values and
clouds include a 1 $\micron$ mean grain size.  The warmer grid-based model
($T_{\mathrm{eff}}$=1800 K) can be ruled out by the larger implied mass ($M_{\mathrm{Jup}}$=40.04) and non-physical radius (0.56 $R_{\mathrm{Jup}}$) inconsistent with
evolutionary models.  Figure \ref{fig:2M0850AB} shows how the grid fit for 2MASS 0850+10B
compares to a cooler model with fixed evolutionary parameters.  We can rule out the warmer
1500 K model because the implied mass (69.91 $M_{\mathrm{Jup}}$) is higher than the total
measured mass of the entire system ($M_{\mathrm{tot}}$ = 54 $\pm$ 8 $M_{\mathrm{Jup}}$).  For the
$T_{\mathrm{eff}}$=1234 K model, the fit can be improved at $J$ band if the grain size is
increased to 1 $\micron$, but this results in a poor fit to HST photometry near 0.75-1
$\micron$.  A deeper cloud can also improve the $J$-band fit but reddens the $H$ and $K$
bands.

We believed the warmer, initial grid fit for this system, notably the primary, may have been caused by the spacing of our grid since evolutionary-predicted gravities for 2MASS 0850+10AB are near the edge of a boundary. To test this hypothesis and determine if the issue was indeed lack of sufficient grid coverage, we extended the segment of our grid that begins at log(g)=4.75 down to log(g)=4.5, including the same range of temperatures and cloud properties given in row two of Table \ref{tab:ModelParams}. We then recalculated the grid fits for both A and B components. The best grid fit for 2MASS 0850+10A did not change with the addition of new lower gravity grid models. The weighted mean properties remained nearly identical to our previous findings and inconsistent with evolutionary predictions. The updated best grid fit for 2MASS 0850+10B resulted in an unusually cool model with a non-physical radius. Weighted mean parameters were again consistent with predictions from evolutionary models within the uncertainties, but the average gravity was still higher than expected (log(g)=5.25$_{+0.25}^{-0.77}$).

Previous work has suggested 2MASS 0850+10A may be an unresolved binary. \citet{Burgasser2011} pointed out the object had unusually bright $J$ and $K$-band absolute magnitudes for a late L dwarf. \citet{Dupuy2012, Dupuy2017} later determined with an updated system distance absolute magnitudes were similar to other L5-L7 dwarfs. Our HST photometry shows large differences in component brightnesses across multiple bands ($\Delta$F625W = 1.31 dex, $\Delta$F775W = 1.16 dex, $\Delta$F814W = 1.47 dex, and $\Delta$F850LP = 0.86 dex) hinting 2MASS 0850+10A might actually be composed of more than one object after all. The small total system mass of 54 $\pm$ 8 $M_{\mathrm{Jup}}$ implies 2MASS 0850+10A would be a pair of objects near the deuterium burning limit ($\approx$ 13.5 $M_{\mathrm{Jup}}$) with lower gravity (log(g)=4.3, $R_{\mathrm{Jup}}$=1.3) assuming the primary is an equal mass binary, and the B component is a more massive single object with higher gravity (log(g)=4.70, $R_{\mathrm{Jup}}$=1.15). The effective temperature in substellar evolutionary models can be quite flat for objects near 20 $M_{\mathrm{Jup}}$ at these ages due to clouds, and low-mass objects of similar effective temperatures have been known to look like mid-to-late L dwarfs \citep{Barman2011}. Fitting the A component as a binary would require more specialized atmosphere models.

\subsection{2MASS 1728+39AB}\label{sec:1728}
\begin{figure*}
\plotone{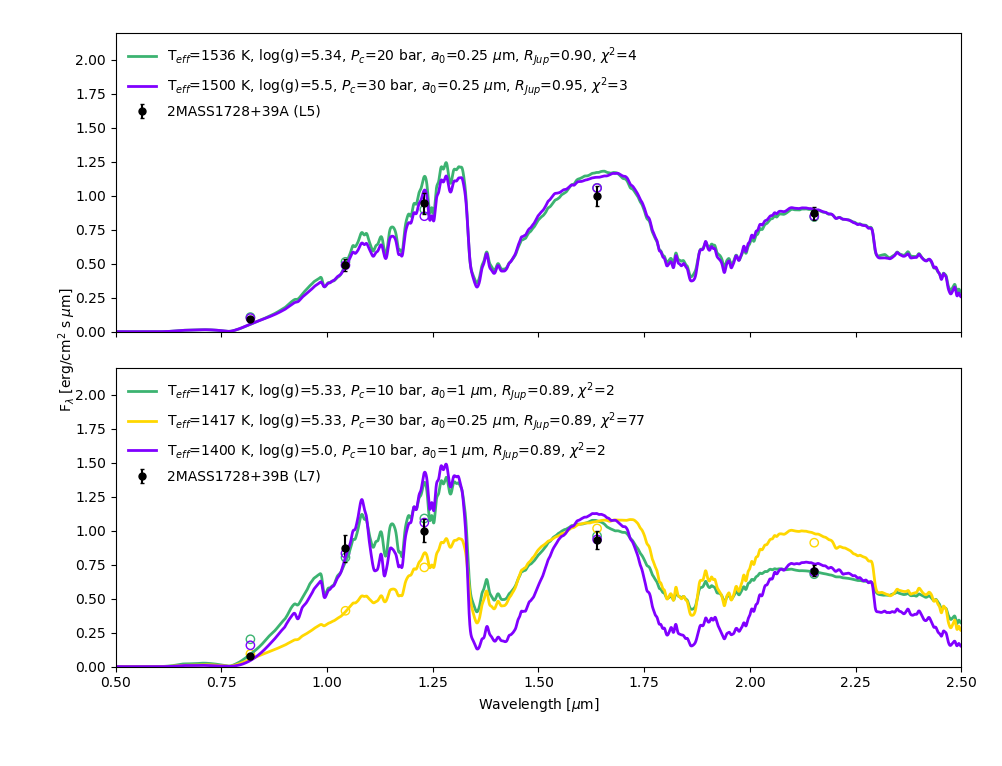}
\caption{New evolutionary fit atmosphere models (green) compared to best grid fits
(purple) for 2MASS 1728+39A (top) and 2MASS 1728+39B (bottom). Two fits for 2MASS 1728+39B
are provided: one excluding the F814W photometric point (green) and one including it
(gold). Surface gravity is in cm $s^{-2}$.\label{fig:2M1728AB}}
\end{figure*}

The best grid fit for 2MASS 1728+39A is consistent with evolutionary predictions in
temperature and radius but gravity is over-predicted resulting in an implied mass that is
inconsistent with measured masses ($M_{\mathrm{Jup}}$=115.22).  By fixing
T$_{\mathrm{eff}}$, log(g), and R$_{\mathrm{Jup}}$ to evolutionary predictions from Table
\ref{tab:Evo} and adjusting the depth of the cloud, we are able to produce a good fit to
the data shown in Figure \ref{fig:2M1728AB}.  The 1536 K model has an implied mass
consistent with measured mass ($M_{\mathrm{Jup}}$=71.54).

Fitting 2MASS 1728+39B was not as straightforward as fitting the A component.  It was
difficult to find a model fit that agreed reasonably well with the data simultaneously in
both the F814W and F1042W bands even with fixed T$_{\mathrm{eff}}$, log(g), and
R$_{\mathrm{Jup}}$ to evolutionary predictions.  The F814W data point was fit best by
models with deeper clouds and 0.25 $\micron$ grains whereas the F1042W point preferred
models with higher clouds and $\geq$ 1 $\micron$ grains.  Figure \ref{fig:2M1728AB} shows
evolutionary model fits with both cloud preferences compared to our initial grid fit. The
lower gravity 1400 K model can be excluded because its implied mass of
$M_{\mathrm{Jup}}$=31.98 is less than half of the measured mass. For the remainder of this
paper, we use the model fit that excludes the F814W photometric point as our preferred fit
because it is more consistent with the majority of the photometric bands.

2MASS 1728+39AB is a flux reversal binary system, which may explain some of the
difficulty finding an atmosphere model that fit in all bands.  Several binary systems have
been discovered with a secondary component brighter than the primary component in the
1.0-1.3 $\micron$ range \citep{Gelino2014}.  In this system, the B component is brighter in the F1042W band ($\Delta_{A-B}$= 0.25 dex) but not in the $J$ or F814W bands.
\citet{Looper2008a} explain the brightening could be the result of a cryptobinary, but it
is often an intrinsic property of the object due to weather, unusual cloud properties, or
changes in surface gravity.  Surface gravity is unlikely to be the culprit because widely
varying gravities requires different ages not expected for coeval binaries
\citep{Gelino2014}.  Our model fits suggest it may be due to differences in cloud location
and particle sizes.

\subsection{2MASS 0920+35AB}
\citet{Dupuy2017} suggest 2MASS 0920+35AB is a triple brown dwarf system given the large
total mass of the system (187 $\pm$ 11 $M_{\mathrm{Jup}}$) with over half of the total
system mass belonging to the fainter secondary (116$^{+7}_{-8}$ $M_{\mathrm{Jup}}$). In
order to obtain evolutionary predictions for the A component of this system, we follow the
same approach as \citet{Dupuy2017} and assume 2MASS 0920+35B is composed of two
equal-mass, equal-luminosity components.  These predictions are given in Table
\ref{tab:Evo}.

\begin{figure*}
\plotone{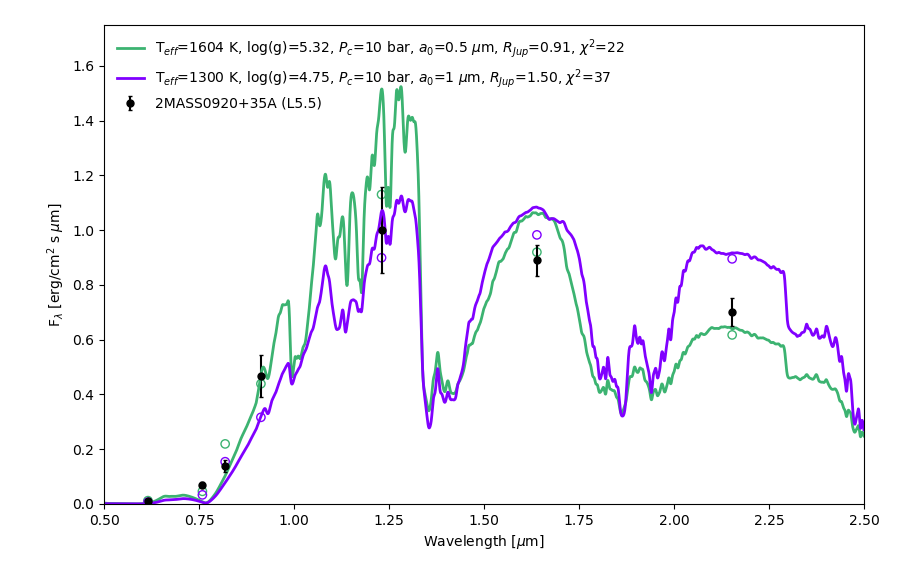}
\caption{New evolutionary fit atmosphere model (green) compared to best grid fit (purple)
for 2MASS 0920+35A. Surface gravity is in $s^{-2}$. \label{fig:2M0920A}}
\end{figure*}

Figure \ref{fig:2M0920A} compares our best grid-fit model of 2MASS 0920+35A from Section
\ref{sec:results} to a new model with fixed T$_{\mathrm{eff}}$, log(g), and
R$_{\mathrm{Jup}}$ from evolutionary predictions.  We are able to produce a good fit to
the data and remain consistent with evolutionary predictions if the B component is indeed
a binary with two equal-mass, equal-luminosity components. We derive a younger age of the
system at 1.82$^{+0.35}_{-0.18}$ Gyr using SM08 evolutionary models, but our results are
consistent within the uncertainties to the system age from \citet{Dupuy2017} of
2.3$^{+0.3}_{-0.4}$ Gyr.

\subsection{LHS2397aAB}
\begin{figure*}
\plotone{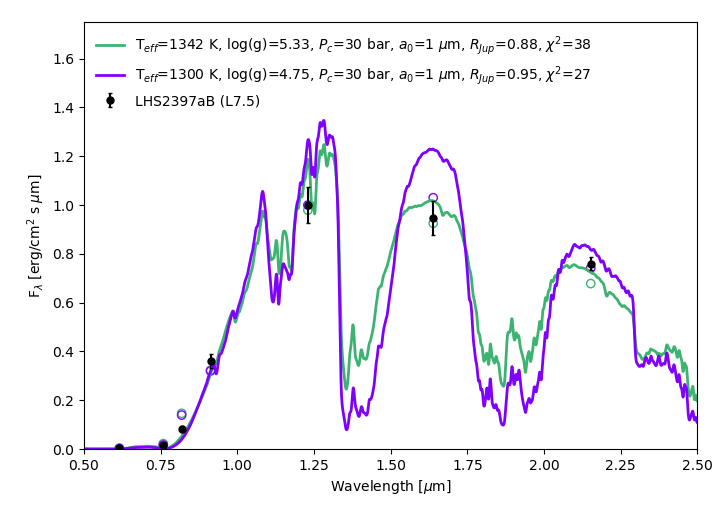}
\caption{New evolutionary fit atmosphere model (green) compared to the best grid fit
(purple) for LHS 2397aB. Surface gravity is in cm $s^{-2}$.\label{fig:LHS2397B}}
\end{figure*}

We only studied the L dwarf companion in this system because the A component is a low-mass
star and outside the temperature range of our grid fitting ($\geq$ 2000 K).  Although the
best grid-fitting model for LHS 2397aB was consistent in temperature and radius to
evolutionary properties, the lower gravity resulted in an implied mass that was too small
($M_{\mathrm{Jup}}$=20.49).  By fixing the bulk atmospheric properties of the object to
evolutionary-derived properties from Table \ref{tab:Evo}, we are able to get a good fit to
the data with a new model show in Figure \ref{fig:LHS2397B} that has an implied mass
consistent with observations ($M_{\mathrm{Jup}}$=66.84).

\subsection{SDSS 1021-03AB}
It became apparent our cloudy atmosphere grids did not sample enough of the parameter
space in cloud pressure and grain size to accommodate early-to-mid T dwarfs, particularly
at lower values of surface gravity near log(g)=4.75. $K$-band flux is sensitive to surface
gravity \citep{Saumon1994}, and the cooler models in our grid more appropriate for T
dwarfs predicted an overly red $H$ - $K$ color.  We were able to obtain improved fits more
appropriate for the SDSS 1021-03AB system by testing models outside the grid with larger
grain sizes and deeper clouds.

\begin{figure*}
\plotone{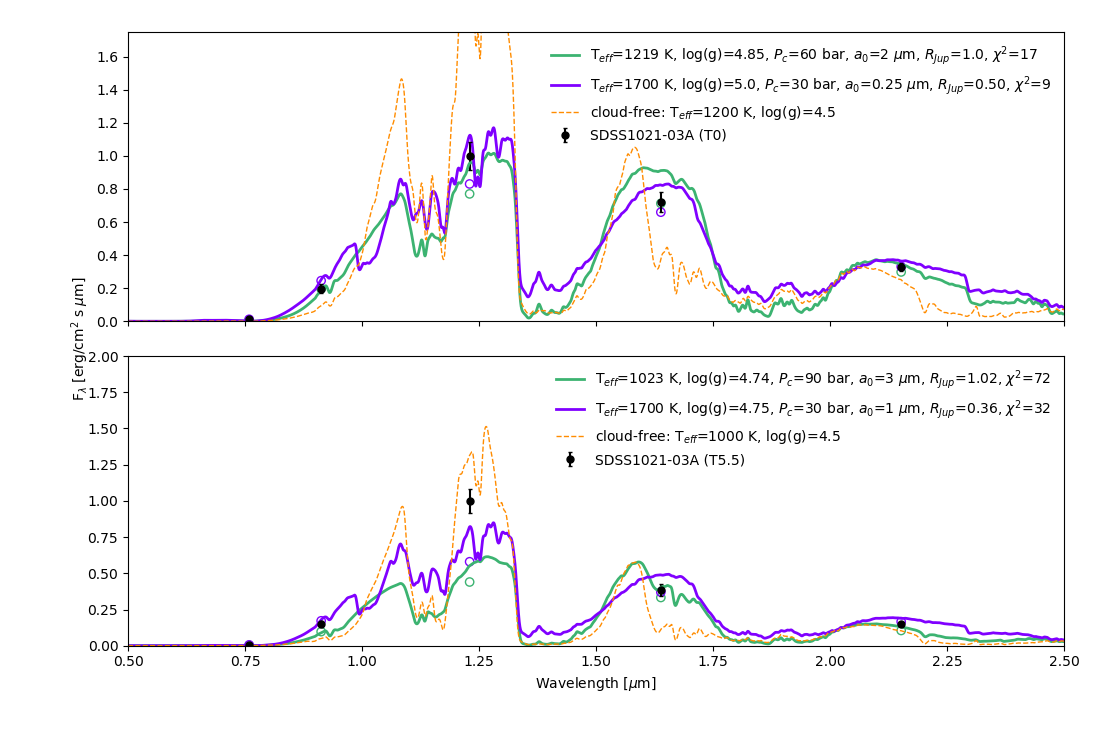}
\caption{New evolutionary fit atmosphere models (green) compared to best grid fits
(purple) for SDSS 1021-03A (top) and SDSS 1021-03B (bottom). Plotted in orange is a
cloud-free model for comparison. Surface gravity is in cm $s^{-2}$.
\label{fig:SDSS1021AB}}
\end{figure*}

The cloudy model fits presented here help to provide an upper limit to the location of
clouds in T dwarf atmospheres, which are traditionally fit with cloud-free models
\citep{Baraffe2003}.  Figure \ref{fig:SDSS1021AB} provides a comparison of the best grid
models to new evolutionary-based models. The initial grid fits to the system have lower
values of  $\chi^{2}$ but significantly over-predicted temperature ($\geq$ 500 K) and
under-predicted radius ($\leq$ 0.50 $R_{\mathrm{Jup}}$). Although the $J$-band flux is
under-predicted by our evolutionary fit for SDSS 1021-03B, this model is still the
preferred model based on mass and radius. If we allow the radius to vary, the fit at
$J$-band can be improved using a model with a deeper cloud and larger grain size
($P_{c}$=100 bar, $a_{0}$=5 $\micron$) with a radius of 1.25 $R_{\mathrm{Jup}}$; however, this
results in a 35 $M_{Jup}$ mass inconsistent with observations of 23 $\pm$ 4.

\subsection{2MASS 1534-29AB} 
Similar to the SDSS 1021-03AB system, the mid-T dwarfs in the 2MASS 1534-29AB system were
not fit well by our cloudy atmosphere grids. Grid fits resulted in significantly
over-predicted effective temperatures ($\geq$ 500 K) and small, non-physical radii ($\leq$
0.40 $R_{\mathrm{Jup}}$) for both the A and B components. 

\begin{figure*}
\plotone{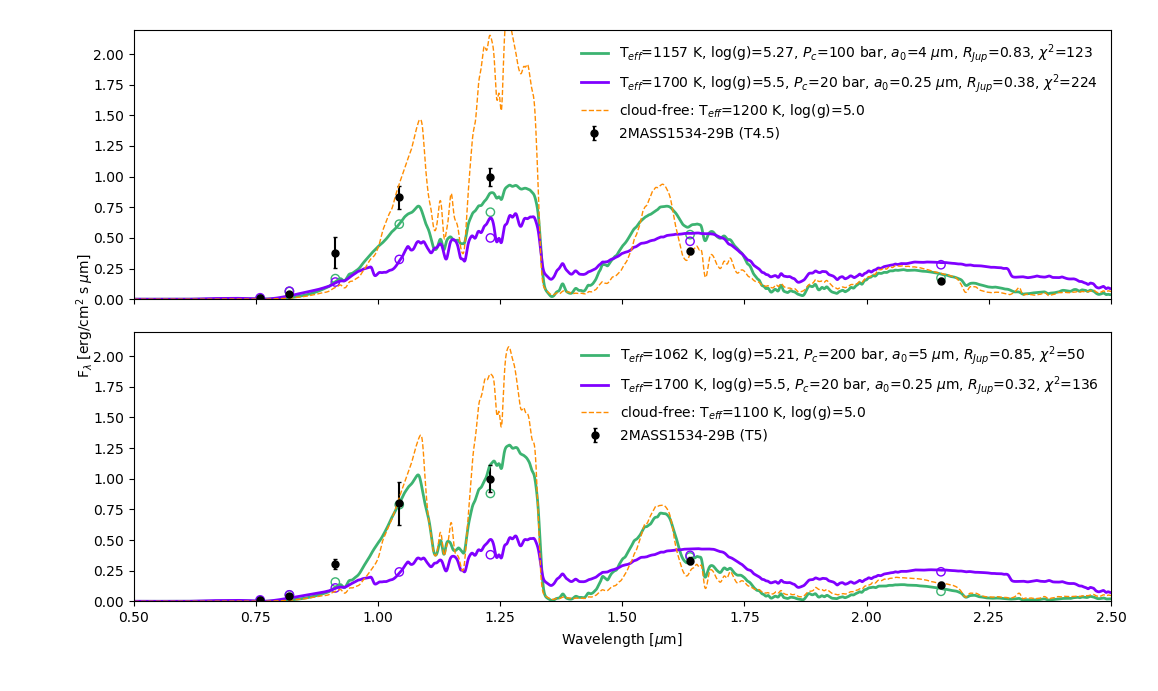}
\caption{New evolutionary fit atmosphere models (green) compared to best grid fits
(purple) for 2MASS 1534-29A (top) and 2MASS 1534-29B (bottom). Plotted in orange is a
cloud-free model for comparison. Surface gravity is in cm $s^{-2}$. \label{fig:2M1534AB}}
\end{figure*}

Using cloud parameters informed from our evolutionary fits to the SDSS 1021-03AB system,
we were able to test additional models and find more realistic fits for 2MASS 1534-29AB
shown in Figure \ref{fig:2M1534AB}. The fits are greatly improved over the original
grid-based models and help to provide upper limits to the cloud over cloud-free models.

\section{Cloud Properties}
Here we discuss detailed cloud properties from our best-fitting atmosphere models.  For
the remainder of the paper, we consider the best fits those constrained by evolutionary
models in the previous section. Model parameters are summarized in Table
\ref{tab:CloudInfo}.  We also include the location of the photosphere, cloud location
(top, base), cloud thickness, and peak gas-to-dust ratio in the table for reference.  We
adopt the approximate location of the spectrum-forming region (photosphere) where the
atmospheric temperature equals the effective temperature. This location is also very
close to where the Rosseland mean optical depth is $\approx$ 1.

The model cloud is composed of multiple layers of thermochemically permissible condensate
species composed mostly of Fe and Mg-Si grains.  Cloud opacity is determined from a
heterogeneous mixture of these grains along with a number of minor contributors for which
opacities are available rather than using a single grain type as representative of all
cloud particles.  We report the most abundant condensate species for each object in Table
\ref{tab:Dust}.  While the table is representative of the composition at the cloud top
layer, a gradient of condensates is present within the cloud.

\begin{figure*}
\includegraphics[width=\textwidth]{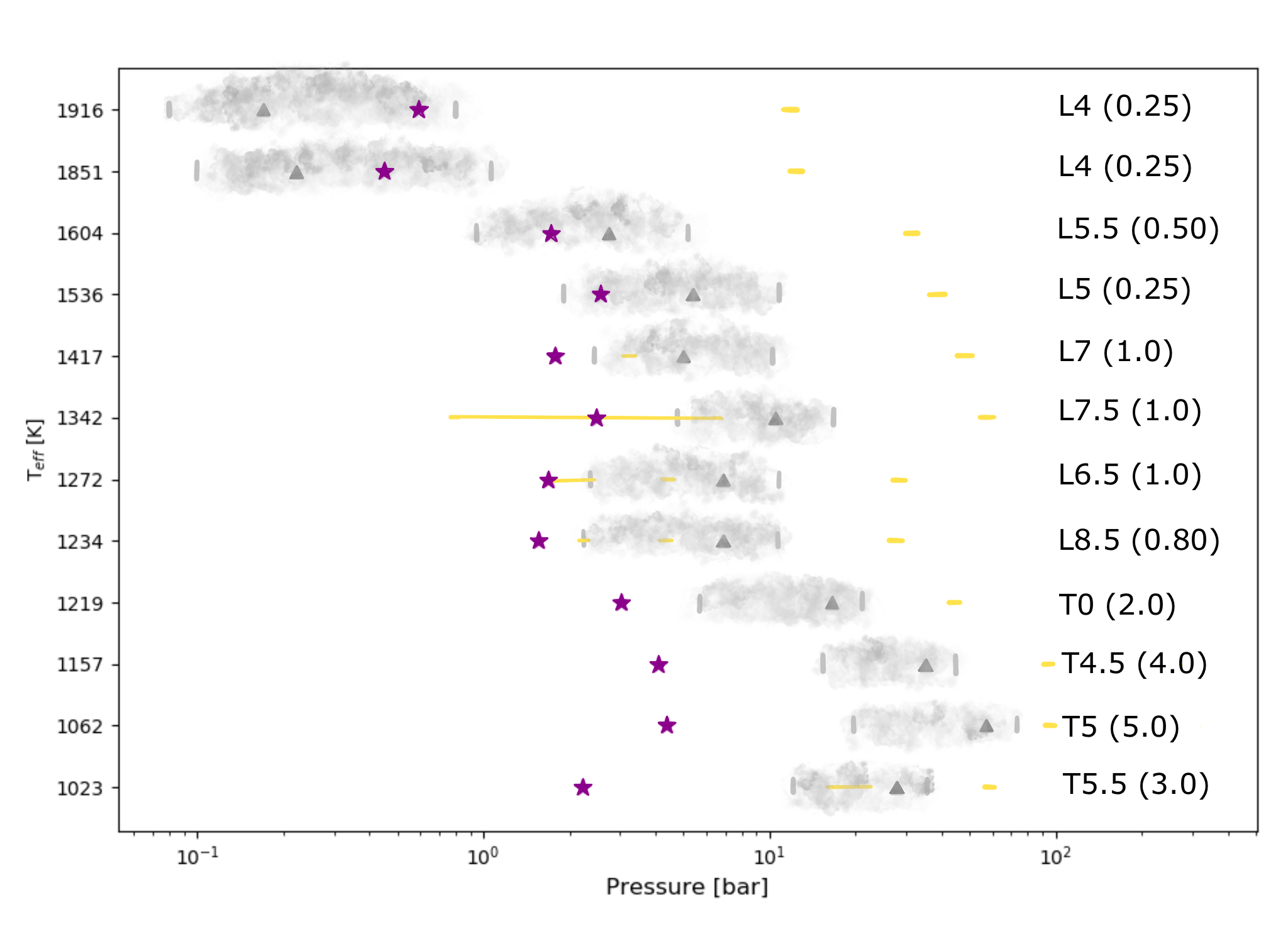}
\caption{Summary of best-fit evolutionary model cloud progression organized by decreasing
effective temperature with the warmest objects at the top. The photosphere for each model
atmosphere is represented by a star.  The cloud top and cloud base are denoted by
light grey lines. The peak dust-to-gas ratio is represented by a grey triangle. Regions
with a yellow line are indicative of convection with the right-most dash always indicating
the radiative-convective boundary.  Yellow regions to the left of the this boundary
highlight a detached convection zone (F$_{\mathrm{conv}}$/F$_{\mathrm{tot}}$ $\geq$ 50
\%). Spectral type and the model's mean grain size in $\micron$ is included on the right-hand side of the plot for reference.\label{fig:CloudPlot}}
\end{figure*}

Figure \ref{fig:CloudPlot} illustrates how the cloud properties change in our sample for L
to mid T-type dwarfs for T$_{\mathrm{eff}} \approx$ 1900-1000 K.  Cloud formation occurs
higher in the atmosphere and shifts to deeper regions as effective temperature decreases.
The peak dust-to-gas ratio shifts to deeper regions within the cloud for cooler objects as
well. This trend emerges into two distinct clusters in the plot--one containing eight
objects ($\approx$ L4-T5) and one containing four objects ($\approx$ L6-T5)--signifying a similar trend
for high-gravity and low-gravity objects. It is worth noting all objects in our sample
near $T_{\mathrm{eff}}$ $\leq$ 1400 K have detached convection zones from the
radiative-convective boundary with a convective flux greater than 50\% of the total flux.
Isolated convection zones can emerge as temperatures decrease.  These detached convection zones are a result of localized changes to the temperature gradient caused by the cloud opacity. Similar zones, produced for the same reason, were discussed by \citet{Burrows2006}. In addition to the effective temperature and gravity, the location and vertical extent of these zones are sensitive to the detailed cloud properties (e.g., composition, particle size, and cloud morphology) that determine a cloud's contribution to the total opacity.

Sub-micron grains (0.25-0.50 $\micron$) were the most appropriate for L4-L5.5 spectral
types with temperatures of $\approx$1900-1500 K and high surface gravities (log(g) $\geq$
5.0) and one lower gravity (log(g) $\leq$ 5.0) L8.5 dwarf.  Previous work found comparable
grain sizes for L dwarfs using a single grain type such as 0.4-0.6 $\micron$ for corundum
and enstatite grains \citep{Marocco2014}, 0.15-0.3 $\micron$ for iron grains
\citep{Marocco2014}, and a mean grain size of 0.15-0.35 $\micron$ using forsterite grains
\citep{Hiranaka2016}.  Rotational modulations have also suggested hazes present in L dwarf
atmospheres have characteristic grain sizes of $\approx$0.28-0.4 $\micron$
\citep{Lew2016}.   

Larger grain sizes ($\geq$ 1 $\micron$) were required for most spectral types later than
L6.5-L7 with effective temperatures near 1400 K and cooler. The coolest and latest type
objects in our sample required the largest grain sizes (2-5 $\micron$). Similarly,
\citet{Zhou2018} found characteristic condensate particle sizes grew for later L types
($\geq$ L8) with larger mean grain sizes required to match observations (a$_{0}$ $\geq$
1.0 \microns).

\begin{figure*}
\includegraphics[width=\textwidth]{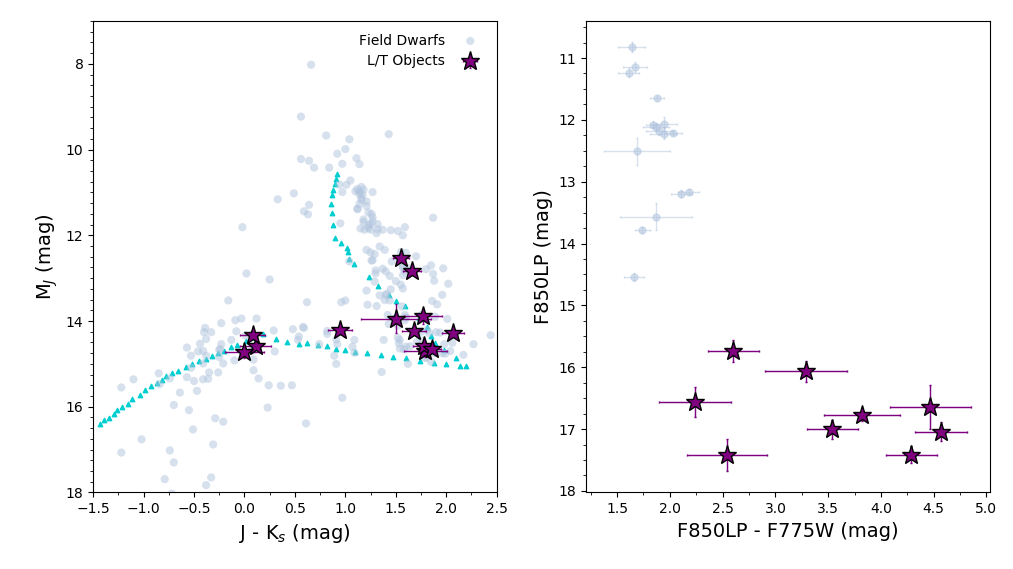}
\caption{Color-magnitude diagrams in near-infrared and optical colors.  \citet{Saumon2008}
hybrid evolutionary tracks are overplotted in turquoise triangles. Purple stars show the
objects focused on in this paper. Filled light blue circles are low-mass field objects
from \citet{Faherty2012} (left) and low-mass objects from \citet{Konopacky2010} (left and
right).  HST observations on right include both WFC3 and ACS filters.}
\label{fig:CMDs}
\end{figure*}

We compare our modelled objects to the cooling and color evolution of brown dwarfs across
the L/T transition using \citet{Saumon2008} hybrid evolution models in Figure
\ref{fig:CMDs}.  Objects redder than the track were best represented by the smallest grain
sizes, whereas objects bluer than the track where those that required the largest grain
sizes.  Reddened L dwarfs in the L5-L7.5 range can be an important indicator of youth in
brown dwarfs, or the redder colors can be indicative of excess dust and clouds in the
photosphere \citep{Kirkpatrick2005,Filippazzo2015,Faherty2012}.  The reddest object in our
sample, 2MASS 1728+39A, is a mid-L dwarf slightly redder than average ($J$ - $K_{s}$ =
2.07), which we believe is due to excess dust and a small cloud particle size.

\begin{deluxetable*}{lllllcccccl}
\tabletypesize{\footnotesize}
\tablecaption{Summary of Evolutionary Constrained Cloud Properties \label{tab:CloudInfo}}
\tablecolumns{11}
\tablewidth{0pt}
\tablehead{
    \colhead{Object}&
    \colhead{T$_{\mathrm{eff}}$}&
    \colhead{log(g)}&
    \colhead{a$_{0}$}&
    \colhead{P$_{\mathrm{c}}$}&
    \colhead{Photosphere}&
    \colhead{Cloud Base}&
    \colhead{Cloud Top}&
    \colhead{Peak Dust}&
    \colhead{H$_{p}$}&
    \colhead{SpT}
}
\startdata
HD 130948B & 1916 & 5.12 & 0.25 & 0.5 & 0.59 & 0.80 & 0.08 & 0.17 & 1.93 & L4 $\pm$ 1 \\ 
HD 130948C & 1851 & 5.10 & 0.25 & 0.5 & 0.45 & 1.07 & 0.10 & 0.22 & 2.58 & L4 $\pm$ 1 \\
2MASS 0920+35A & 1604 & 5.32 & 0.50 & 10 & 1.72 & 5.21 & 0.94 & 2.74 & 2.14 & L5.5 $\pm$ 1 \\ 
2MASS 1728+39A & 1536 & 5.34 & 0.25 & 20 & 2.57 & 10.8 & 1.89 & 5.37 & 2.16 & L5 $\pm$ 1 \\ 
2MASS 1728+39B & 1417 & 5.33 & 1.0 & 10 & 1.78 & 10.2 & 2.45 & 5.00 & 1.90 & L7 $\pm$ 1 \\
LHS2397aB & 1342 & 5.33 & 1.0 & 30 & 2.49 & 16.7 & 4.79 & 10.5 & 1.53 & L7.5 $\pm$ 1 \\
2MASS 0850+10A & 1272 & 4.72 & 1.0 & 30 & 1.68 & 10.8 & 2.37 & 6.90 & 1.84 & L6.5 $\pm$ 1 \\
2MASS 0850+10B & 1234 & 4.70 & 0.8 & 20 & 1.56 & 10.7 & 2.24 & 6.84 & 1.90 & L8.5 $\pm$ 1  \\
SDSS 1021-03A & 1219 & 4.85 & 2.0 & 60 & 3.04 & 21.0 & 5.67 & 16.5 & 1.57 & T0 $\pm$ 1 \\
2MASS 1534-29A & 1157 & 5.27 & 4.0 & 100 & 4.06 & 44.6 & 15.4 & 34.9 & 1.30 & T4.5 $\pm$ 1 \\
2MASS 1534-29B & 1062 & 5.21 & 5.0 & 200 & 4.35 & 73.6 & 19.6 & 57.2 & 1.48 & T5 $\pm$ 1 \\
SDSS 1021-03B & 1023 & 4.74 & 3.0 & 90 & 2.21 & 35.4 & 12.1 & 27.8 & 1.26 & T5.5 $\pm$ 1 \\
\enddata
\tablecomments{Detailed model cloud properties for best evolutionary-derived atmosphere
model fits from Section \ref{EvoComparison}. Cloud properties are given in bars. H$_{p}$
is the cloud thickness in pressure scale height. Spectral types listed are from
\citet{Dupuy2017}.}
\end{deluxetable*}

\begin{deluxetable*}{ll}
\tabletypesize{\footnotesize}
\tablecaption{Most Abundant Condensates in Cloud Top Layer \label{tab:Dust}}
\tablecolumns{2}
\tablewidth{0pt}
\tablehead{
    \colhead{Object}&
    \colhead{Condensates}
}
\startdata
HD 130948B & Fe (48\%) | MgSiO$_{3}$, Mg$_{2}$SiO$_{4}$, MgO, SiO$_{2}$ (52\%)\\
HD 130948C & Fe (49\%) | MgSiO$_{3}$, MgO, Mg$_{2}$SiO$_{4}$, SiO$_{2}$ (51\%) \\
2MASS 0920+35A & Fe (48\%) |  MgSiO$_{3}$, Mg$_{2}$SiO$_{4}$, MgO, SiO$_{2}$ (52\%)\\
2MASS 1728+39A & Fe (49\%) | MgSiO$_{3}$, MgO, Mg$_{2}$SiO$_{4}$, SiO$_{2}$ (51\%) \\
2MASS 1728+39B & Fe (50\%) |  MgO, SiO$_{2}$, Mg$_{2}$SiO$_{4}$, MgSiO$_{3}$, MgAl$_{2}$O$_{4}$ (50\%)\\
LHS2397aB & Fe (45\%) |  MgSiO$_{3}$, MgO, SiO$_{2}$, Mg$_{2}$SiO$_{4}$ (55\%)\\
2MASS 0850+10A & Fe (49\%) | MgSiO$_{3}$, MgO, Mg$_{2}$SiO$_{4}$, SiO$_{2}$ (51\%) \\
2MASS 0850+10B & Fe (48\%) |  MgSiO$_{3}$, Mg$_{2}$SiO$_{4}$, MgO, SiO$_{2}$ (52\%)\\
SDSS 1021-03A & Fe (47\%) |  MgSiO$_{3}$, MgO, Mg$_{2}$SiO$_{4}$, SiO$_{2}$ (53\%)\\
2MASS 1534-29A & Fe (45\%) | SiO$_{2}$, MgO, Mg$_{2}$SiO$_{4}$, MgSiO$_{3}$, MgAl$_{2}$O$_{4}$ (55\%) \\
2MASS 1534-29B & Fe (46\%) | SiO$_{2}$, MgO, Mg$_{2}$SiO$_{4}$, MgSiO$_{3}$, MgAl$_{2}$O$_{4}$ (54\%) \\
SDSS 1021-03B & Fe (47\%) | MgSiO$_{3}$, MgO, SiO$_{2}$, Mg$_{2}$SiO$_{4}$ (53\%) \\
\enddata
\tablecomments{The most abundant dust species are listed for each object at the cloud top
layer. Individual condensates species are in order from the most abundant to the least
abundant and represent a combination of both solid and liquid phases. Percentages shown
are for total iron grains and Mg-Si grains, respectively.}
\end{deluxetable*}

\section{Summary and Conclusion}
Discrepancies between atmosphere and evolutionary model predictions play a significant
role as different stages of condensate cloud evolution greatly influences the overall
spectral shape.  Previous grid comparisons have led to disagreements between atmosphere
models and evolutionary predictions for substellar objects of known mass
\citep[e.g.,][]{Chabrier2000, Dupuy2014}.  We are able to produce atmosphere models that
match evolutionary predictions for a sample of brown dwarfs ($\approx$ L4-T5) by allowing
enough flexibility within the cloud properties.

Determining the bulk properties (e.g., effective temperature and gravity) of stellar and
substellar mass objects is an important step along the way to making inferences about more
specific properties, such as metallicity, or the relative abundances of key elements.
These bulk properties are frequently estimated by comparing photometric or spectroscopic
observations to model atmosphere predictions. For many objects such comparisons yield
properties consistent with those of interior and evolution models that, at moderate to old
ages, are considered reliable and less sensitive to model assumptions \citep{Baraffe2002}.
Significant inconsistencies, however, can occur whenever condensate cloud formation
greatly influences the overall spectral shape. This situation occurs, for example, across
the L/T transition \citep{Cushing2006} and for most young directly imaged companions
\citep{Metchev2006} where models with a range of temperatures, gravities, and cloud
properties are capable of matching a single object's near-IR SED equally well
\citep{Barman2011}. Such ambiguity can often hinder the study of cloudy objects where mass
and age are very uncertain.

In this paper we have studied a set of L/T transition ($T_{\mathrm{eff}}$ $\approx$ 1900-1000 K) brown dwarf
binaries with measured masses, luminosities, and well-determined ages.  For these objects
comparisons to evolutionary models yield very precise estimates of the bulk properties. Overall, synthetic spectra from our cloudy atmosphere models matched spatially-resolved visible to near-IR photometry of each binary component reasonably well. The cloud parameters included in our grids appear to be most appropriate for L4-L8 field dwarfs with effective temperatures between 1900-1300 K and log(g) $\geq$ 5.0. The grid fits for these objects were the most consistent with our evolutionary-constrained atmosphere models.

With these atmosphere models we have determined a set of cloud properties across the L/T
transition. The warmest objects in the sample ($\approx$1900-1500 K) were fit best by
particles with mean grain sizes of 0.25-0.50 $\micron$, whereas objects cooler than 1500 K
required larger mean grain sizes 0.80-5 $\micron$.  Although the composition at the cloud
top remained relatively close to an equal split between Fe and Mg-Si grains for the
majority of objects, the overall location of the cloud top shifted to deeper regions
within the atmosphere as objects cooled in effective temperature. Near 1400 K clouds began
to disappear below the photosphere, which agrees with previous findings \citep{Saumon2008,Marley2010}.

There was some disagreement between our grid-based atmosphere models and the evolutionary-constrained atmosphere models for lower-gravity objects (log(g) $\leq$ 5.0) and the latest T dwarfs included in the sample (T4.5-T5.5; T$_{\mathrm{eff}}$ $\leq$ 1200 K). Our grid-based models tended to over-predict temperature, gravity, and under-predict radius. In addition to model-grid aspects (e.g., grid spacing and boundaries), unresolved binarity might play a role in some systems. For example 2MASS 0850+10A may be a pair of objects near the deuterium burning limit, rather than a single L-type brown dwarf, due to the large difference in brightness between A and B components, low total mass of the system, and mixed results from atmosphere model fitting. Giant exoplanets overlap with the range of effective temperatures of brown dwarfs \citep{Faherty2016} and such objects (e.g., HR8799c) can resemble mid-L spectral types \citep{Marois2010,Barman2011}. Binarity is just one possibility and further study, both observationally and modeling, is warranted for this system.

SDSS 1021-03A is the earliest T dwarf in the sample, and evolutionary-constrained models preferred a deeper cloud (P$_{c}$=60 bar) and larger grain size ($>$ 1 $\micron$) slightly beyond the parameters in our atmosphere grids. Similarly, the later T4-T5 dwarfs (SDSS 1021-03B, 2MASS 1534-29A, 2MASS 1534-29B) required even deeper clouds (P$_{c}$=90-200 bar) and the largest grain sizes (3-5 $\micron$). Evolutionary-derived fits for these objects suggest our grids should be extended to include a larger range of cloud properties to accommodate early-to-mid T dwarfs and help constrain the limits of homogeneous cloud models for the coolest objects. We observe that condensate growth becomes more important near late-L types, leading to preferred model fits with larger mean grain sizes for T dwarfs. Other work has successfully reproduced T dwarf photometry using thin sulfide clouds \citep{Morley2012} and inhomogeneous cloud cover with low temperature condensates (Na$_{2}$S, KCl) \citep{Charnay2018}. Near T$_{\mathrm{eff}}$ $\approx$ 1000 K cloud-free models may be a better fit to our data; however, this is the same regime in low-gravity objects where sulfide clouds appear while iron and silicate clouds are simultaneous disappearing \citep{Morley2012,Charnay2018}. A more diverse grid for these objects will be beneficial in order to untangle the relationship between condensate growth, cloud composition, and surface gravity.

Rapid color changes across the L/T transition have been interpreted as the result of
patchy clouds or holes in the cloud deck \citep{Burrows2003,Ackerman2001, Marley2010}, a
sudden collapse of the cloud deck \citep{Tsuji2003}, or an increase in sedimentation
efficiency of clouds \citep{Knapp2004}.  By parameterizing the sedimentation efficiency of
dust particles to regulate the influence of cloud opacity on the model spectrum, one can
reproduce the L/T transition \citep{Saumon2008,Stephens2009}. We are able to reproduce
photometric changes across the L/T transition in a similar fashion by parameterizing the
vertical extent and mean grain size of a uniform cloud. However, we cannot rule out the
existence of patchy clouds in this sample of objects despite our homogeneous cloudy model
fits because the presence of cloud holes is very subtle across the near-infrared part of
the spectrum for L/T transition objects \citep{Marley2010, Apai2013}.

Other groups use a more complex treatment of grain nucleation, growth, evaporation, and/or drift. Work from \citet{Helling2008,Helling2008b} resulted in mean cloud particle sizes that increased as a function of atmospheric depth, with small particles ($\approx$ 0.01 $\micron$) in high atmospheric layers with a narrow grain size distribution that broadened to larger particle sizes ($\approx$ 100 $\micron$) and grain size distributions near the cloud base. \citet{Charnay2018} was able to reproduce the spread of near-IR colors across the L/T transition and those observed in reddened low-gravity objects by computing cloud particle radii estimated from simple microphysics. Our models use a constant grain size distribution with cloud height, but similar to more complex treatment of grains, the peak dust-to-gas ratio sinks to deeper regions within the atmosphere, eventually below observable layers.

It has been hinted at that grain size increases as objects cool across the L/T transition
\citep{Zhou2018,Knapp2004}. \citet{Burrows2006} used models with homogeneous forsterite
grains with sizes of 3, 10, 30, and 100 $\micron$ and identified that atmospheres with
larger particles resulted in stronger $J$-band fluxes near 1400-1500 K due to the natural
deepening of the cloud position in the model atmospheres.  We observe something similar in
our grid models across a smaller range of particle sizes (0.25-1 $\micron$) for our
heterogeneous grains.  Figure \ref{fig:GSPlot} shows $J$-band flux is greater for the
largest grain sizes at 1300 K whereas $J$-band fluxes are similar at 1500 K regardless of
grain size. A physical mechanism responsible for this trend in increasing grain size is
not well understood.  At cooler effective temperatures, it has been suggested that a
larger supply of condensate vapor near the base of the cloud could result in runaway
particle growth for cooler objects \citep{Gao2018}.

Photometry has limited sensitivity; therefore, future steps will be to improve atmospheric
constraints with resolved spectroscopy.  Cloud location, mean grain size, surface gravity,
and metallicity impact our understanding substellar atmospheres as a function of
temperature.  Decoupling the degeneracies between these interwoven features is essential
to explain observations of objects, particularly across the L/T transition.  This work
provides valuable insight into the complex evolution of cloud opacity for a range of
well-studied brown dwarf binaries. Ideally, the goal is to be able to construct reliable
atmosphere models that can account for the drastic color change and physical properties of
substellar objects without a reliance on evolutionary models to infer the properties of
individual objects.  We plan to extend our cloudy grids to both lower and higher cloud top
pressures with additional grain sizes to accommodate the earliest and latest spectral
types.  Additional exploration of cloud properties is warranted for the warmest and
coolest objects for binaries where our atmosphere grids lacked coverage.

Future work would greatly benefit from additional observations, especially at the shortest
wavelength portion of the SED.  Photometric bands near 0.8-1 $\micron$ appear important
when investigating flux reversal binary systems and provide insight to differing grain
size preferences at optical and near-infrared wavelengths.  The next generation of
telescopes with higher sensitivity and wavelength coverage will be essential in providing
high-quality spectra required to fine tune cloud parameters and address lingering
inconsistencies, such as broad, continuous wavelength coverage from the James Webb Space Telescope's NIRSpec instrument (0.6-5.3 $\micron$). Furthermore, variability has been detected in brown dwarfs at the
transition region \citep{Radigan2014} and will be an important factor for cloud formation
and evolution going forward.  Rotation-modulated spectral variations will be a key
approach toward a more in-depth grasp of the evolving cloud structure in low-mass objects
\citep{Apai2013,Apai2017}.  Understanding the relationship between grain size distribution
and effective temperature will require a multidimensional approach to grain kinetics and
growth connected to convective cloud structure. Atmospheric retrieval results can be compared to those of self-consistent models to better understand the nature of discrepancies (e.g., heterogeneous cloud layers, particle sizes, cloud composition, haze layers). Relatively few brown dwarfs spanning the L/T transition have known masses, and the release of future \textit{Gaia} data will increase parallax precision by 30\% \citep{GAIA2020}. Objects with independently constrained properties from dynamical mass and luminosity measurements are the strongest candidates for future model comparisons.

\section{Acknowledgements}
L. S. B. is extremely grateful for Kyle Pearson and many helpful discussions of this work. We would also like to thank the anonymous referee for a constructive report.
Support for program 11605 was provided by NASA through a grant from the Space Telescope
Science Institute, which is operated by the Associations of Universities for Research in
Astronomy, incorporated under NASA contract NAS5-26555. This work was also supported by
NSF grants 1405505 and 1614492. Material presented in this work is supported by the
National Aeronautics and Space Administration under Grants/Contracts/Agreements
No.NNX17AB63G issued through the Astrophysics Division of the Science Mission Directorate.
This research makes use of data products from 2MASS, a joint project of the University of
Massachusetts and the Infrared Processing and Analysis Center/California Institute of
Technology, funded by the National Aeronautics and Space Administration and the National
Science Foundation. Some of the data presented herein were also obtained at the W. M. Keck
Observatory, which is operated as a scientific partnership among the California Institute
of Technology, the University of California, and the National Aeronautics and Space
Administration. The Observatory was made possible by the generous financial support of the
W. M. Keck Foundation. The authors also recognize and acknowledge the very significant
cultural role and reverence that the summit of Mauna Kea has always had within the
indigenous Hawaiian community.  We are most fortunate to have the opportunity to conduct
observations from this mountain. Additionally, this research has benefited from the SpeX
Prism Library, maintained by Adam Burgasser at http://www.browndwarfs.org/spexprism.

\software{TinyTim \citep{Krist2011}, StarFinder \citep{Diolaiti2000}, img2xym \citep{Anderson2006}, PHOENIX \citep{Hauschildt1993,Baron2006,Baron2007}}

\bibliography{main}{}
\bibliographystyle{aasjournal}

\end{document}